\documentclass[pdflatex]{revtex4}

\usepackage{latexsym}
\usepackage{amsmath,amssymb,bm} 
\usepackage[pdftex]{graphicx} 
\usepackage[pdftex, colorlinks=true, linkcolor=blue, citecolor=blue]{hyperref} 
\usepackage{comment,ulem}

\makeatletter

\newcommand{\cket}[1]{| #1 \rangle} 
\newcommand{\bra}[1]{\langle #1 |} 
\newcommand{\ImUnit}{\mathrm{i}} 
\newcommand{\evec}[1]{\boldsymbol{#1}} 

\begin{document}

\title{Entanglement extracted from vacuum into accelerated Unruh-DeWitt detectors 
and energy conservation}

\author{Jun-ichirou Koga} 
\affiliation{Research Institute for Science and Engineering, 
Waseda University, Shinjuku-ku, Tokyo 169-8555, Japan} 
\email{koga@waseda.jp} 

\author{Kengo Maeda} 
\affiliation{Faculty of Engineering, 
Shibaura Institute of Technology, Saitama 330-8570, Japan} 
\email{maeda302@sic.shibaura-it.ac.jp}
 
\author{Gen Kimura} 
\affiliation{College of Systems Engineering and Science, 
Shibaura Institute of Technology, Saitama 330-8570, Japan} 
\email{gen@shibaura-it.ac.jp}

\begin{abstract} 
We consider a pair of two-level Unruh-DeWitt detectors accelerated uniformly in the Minkowski vacuum of a massless neutral scalar field, 
and analyze, within the perturbation theory, the entanglement extracted from the vacuum into the Unruh-DeWitt detectors when the switching of the detectors are performed adiabatically enough at the asymptotic past and future. We consider the cases where 
the detectors are accelerated parallelly, anti-parallelly, and in differently orientated directions. 
We show that entanglement is extracted if they are accelerated anti-parallelly and the ratios of the excitation energy to 
the magnitude of the acceleration coincide between the two detectors. 
On the other hand, we find the detectors are not entangled when the detectors are accelerated parallelly or in orientated directions. 
We discuss these results from the viewpoint of the energy conservation associated with the timelike boost Killing vector fields tangent to 
the worldlines of the detectors. 
\end{abstract}

\maketitle 

\section{Introduction}  

The discovery that a vacuum contains entanglement \cite{SummersWerner} has stimulated various researches 
on extraction of entanglement from a vacuum of a quantum field into a pair of quantum detectors in a flat spacetime, 
where it is supposed to define operationally the entanglement contained in the vacuum. (See, e.g., Refs.  \cite{Reznik03,SaltonMM15,LInCH08,LinHu10,LinCH15,MartinMartinezST16,AhmadiLCSMD16,GrochowskiLD19-,HerdersonHMSZ17,KogaKM18} and references therein.)  
Although each detector and a quantum field are entangled naturally due to the coupling between them, 
why and how a pair of detectors are entangled looks non-trivial, particularly in vacuum and when spatially separated. 
Investigations into this issue are thus expected to unveil the nature of the entanglement in a vacuum. 

In addition, effects of gravity on entanglement will be crucial both experimentally and theoretically. 
In particular, understanding the behavior of quantum information under the effects of gravity seems inevitable in order to 
resolve the information paradox of black hole \cite{Hawking76}. 
From the viewpoint of the equivalence principle, whether or not it remains valid in the quantum regime, as in the classical general relativity, 
it is an important step to investigate the effects of acceleration on quantum information. 

Indeed, many papers have appeared which have investigated effects of acceleration on entanglement extraction from a vacuum 
into a pair of detectors, 
which include the investigations using two-level detectors \cite{Reznik03,SaltonMM15}, 
those using detectors of the type of harmonic oscillator \cite{LInCH08,LinHu10,LinCH15}, 
and the analyses based on wave packets \cite{AhmadiLCSMD16,GrochowskiLD19-}. 
Extension to a black hole spacetime has been also considered \cite{HerdersonHMSZ17}. 
However, the models considered and the results derived in these investigations vary among them. 
Depending on models under consideration, detectors are found to be entangled in some cases, but not entangled in other cases. 
It does not seem to be clearly understood what sorts of elementary physical processes work behind the various phenomena of 
entanglement extraction. 

One of the most important keys to understanding the effects of acceleration on quantum phenomena will undoubtedly be 
Unruh effect \cite{Unruh76}. 
On one hand, one may naturally expect that accelerated detectors in a flat spacetime will be entangled because of Unruh effect, 
as it has been actually demonstrated in Ref. \cite{Reznik03}, 
since the essential ingredient in Unruh effect is the fact that the left and right Rindler wedges are strongly correlated.  
On the other hand, it is also conceivable that thermal fluctuations due to Unruh effect will force detectors to decohere, 
and thus entanglement will be degraded, as shown in Refs. \cite{LInCH08,LinHu10,LinCH15}. 
In addition to Unruh effect, more intricate motions, spatial extension and structures, and switching effects of detectors may complicate the behavior of entanglement extraction. 

In the previous paper \cite{KogaKM18}, we analyzed entanglement extracted from the Minkowski vacuum into a pair of inertial Unruh-DeWitt detectors \cite{Unruh76,DeWitt79,BirrellDavies82}, and found, within the perturbation theory, but for the general monopole coupling, that the detectors prepared in the ground state at the asymptotic past are not entangled at the asymptotic future, 
if they are comoving and they are switched on and off adiabatically enough at the asymptotic past and future. 
This result has thus provided a fiducial system, based on which one can argue various physical effects on entanglement extraction. 
In particular, we also found that when the detectors are in a relative inertial motion, entanglement is extracted due to the special relativistic effect. 
We recall here that the switching considered in \cite{KogaKM18} was adopted in order for Unruh-DeWitt detectors to probe faithfully the feature of the Minkowski vacuum, i.e., as a true vacuum (no particles) for an inertial observer, 
but as a thermal state for a uniformly accelerated observer. 

The result that no entanglement is extracted in the case of the fiducial system as above, i.e., comoving inertial Unruh-DeWitt detectors 
in the Minkowski vacuum, was found to be described by the delta function representing the energy conservation \cite{KogaKM18}. 
In this paper, we will extend this to the case of uniformly accelerated Unruh-DeWitt detectors. 
Since it would not be possible to treat the delta function numerically, it is advantageous to consider analytically tractable cases to 
analyze energy conservation. 
Among them, 
we will consider in this paper  a pair of two-level Unruh-DeWitt detectors in the Minkowski vacuum of a massless neutral scalar field,  
which are uniformly accelerated parallelly, anti-parallelly, or in differently orientated directions with each other. 

We will prepare in Sec. \ref{sec:model} the model we will consider, the same model as in Ref. \cite{KogaKM18}, 
and review entanglement measures. 
In Sec. \ref{sec:entangle}, after presenting our framework, we will analyze entanglement extraction in each of parallel acceleration, 
anti-parallel acceleration, and acceleration into orientated directions, with or without a translational shift of the worldlines of the detectors, 
as well as different magnitudes of accelerations. 
Sec. \ref{sec:summary} is devoted to the summary of this paper and discussion. In Appendix, by applying the framework presented in this paper, 
we will consider uniformly accelerated detectors that reduce to those in a relative inertial motion in the vanishing acceleration limit, 
and we will see that it reproduces the result in Ref. \cite{KogaKM18}. 
Natural units $c = \hbar = k_B = 1$ are used throughout this paper.

\section{Model} 
\label{sec:model} 

We consider a pair of two-level Unruh-DeWitt detectors, $A$ carried by Alice and $B$ by Bob, which are considered as two qubits, 
in the Minkowski vacuum of a massless neutral scalar field $\phi(x)$ in a flat spacetime. 
The energy eigenstates and the corresponding eigenvalues of the detector $I$, where $I$ stands for $A$ or $B$, are defined with respect to the proper time $\tau_I$ of the detector $I$, and denoted as 
$\cket{E_n^{(I)}}$ and $E_n^{(I)}$, respectively,
where $n = 0 , 1$. We assume in this paper that the energy eigenstates are not degenerate, 
and hence the excitation energy of each detector is greater than zero, 
\begin{equation} 
\Delta E^{(I)} \equiv E_1^{(I)} - E_0^{(I)} > 0 . 
\label{eqn:NonDegenerate} 
\end{equation} 
The coupling of these Unruh-DeWitt detectors with the scalar field $\phi(x)$ is described by 
the interaction action 
\begin{equation}
\mathcal{S}_{{\rm int}} 
= \int c \, \chi_A(\tau_A) \, m_A(\tau_A) \, \phi(\bar{x}_A) d \tau_A 
+ \int c \, \chi_B(\tau_B) \, m_B(\tau_B) \, \phi(\bar{x}_B) d \tau_B , 
\label{eqn:IntAction} 
\end{equation}
where $c$ is the sufficiently small coupling constant, 
$\bar{x}_I^{\mu}(\tau_I)$ denotes the worldline coordinates of the detector $I$, and 
$m_I(\tau_I)$ is an arbitrary monopole operator of the detector $I$, which commutes with that of the other detector and with the scalar field 
$\phi(\bar{x}_I(\tau_I))$. 
We thus consider the general coupling of a point-like detector, without focusing on a particular form 
as in Ref. \cite{SaltonMM15}. 
The switching function $\chi_I(\tau_I)$ describes how the switching of the detectors is executed. 
In this paper, we exclusively consider the case where the switching is performed adiabatically enough at the asymptotic past and 
future, in contrast to the Gaussian switching function in Ref. \cite{SaltonMM15}, and thus we set as $\chi_I(\tau_I) = 1$ for both the detectors, as in the textbooks \cite{DeWitt79,BirrellDavies82} on Unruh-DeWitt detectors. 

The quantum state of the whole system at the asymptotic past ($t = - \infty$), $\cket{\mathrm{in}}$, 
is assumed to be the product state as  
\begin{equation}
\cket{\mathrm{in}} = \cket{0} \cket{E_0^{(A)}} \cket{E_0^{(B)}}
\end{equation} 
where $\cket{0}$ is the Minkowski vacuum of the quantum scalar field $\phi(x)$. 
Then, the quantum state $\rho_{AB}$ of the two detectors $A$ and $B$ at the asymptotic future ($t = \infty$) is derived based on the standard perturbation theory, and given,  
by tracing out the degrees of freedom of the scalar field, as 
\begin{equation} 
\rho_{AB} = \begin{pmatrix} 
0 & 0 & 0 & c^2 \: \mathcal{E} \\ 
0 & c^2 \: \mathcal{P}_A & c^2 \: \mathcal{P}_{AB} & c^2 \: \mathcal{W}_A \\ 
0 & c^2 \: \mathcal{P}_{AB}^* & c^2 \: \mathcal{P}_B & c^2 \: \mathcal{W}_B \\ 
c^2 \: \mathcal{E}^* & c^2 \: \mathcal{W}_A^* & c^2 \: \mathcal{W}_B^* & 
1 - c^2 \big( \mathcal{P}_A + \mathcal{P}_B \big) \end{pmatrix} + \mathcal{O}(c^4) , 
\label{eqn:RhoAB} 
\end{equation} 
in the basis $\left\{ \cket{E^{(A)}_1} \cket{E^{(B)}_1} , \; 
\cket{E^{(A)}_1} \cket{E^{(B)}_0} , \; \cket{E^{(A)}_0} \cket{E^{(B)}_1} , \; \cket{E^{(A)}_0} \cket{E^{(B)}_0} \right\}$. 
The relevant components of Eq. (\ref{eqn:RhoAB}) in this paper are written as 
\begin{align} & 
\mathcal{P}_I = 
\left| \bra{E_{1}^{(I)}} \, m_I(0) \, \cket{E_0^{(I)}} \right|^2 \, \mathcal{I}_I , 
\label{eqn:CalPADef} 
\\ & 
\mathcal{E} = \bra{E_{1}^{(B)}} m_B(0) \cket{E_0^{(B)}} \, \bra{E_{1}^{(A)}} m_A(0) \cket{E_0^{(A)}} \, \mathcal{I}_E , 
\label{eqn:CalEDefApp1} 
\end{align} 
where 
\begin{align} & 
\mathcal{I}_I \equiv \int_{- \infty}^{\infty} d \tau'_I \; \int_{- \infty}^{\infty} d \tau_I \; 
e^{\ImUnit \, \Delta E^{(I)} \left( \tau_I-  \tau'_I \right)} \: G_W(\bar{x}'_I , \bar{x}_I) , 
\label{eqn:CalIIDef} 
\\ & 
\mathcal{I}_E \equiv
- \: \ImUnit \:  \int_{- \infty}^{\infty} d \tau_B \, \int_{- \infty}^{\infty} d \tau_A \:  
e^{\ImUnit \, \Delta E^{(B)} \tau_B} e^{\ImUnit \, \Delta E^{(A)} \tau_A} \: G_F(\bar{x}_B , \bar{x}_A) , 
\label{eqn:CalETotDef} 
\end{align}
and $G_W(x, x')$ and $G_F(x, x')$ are the Wightman function and the Feynman propagator, 
respectively, which are given for the massless neutral scalar field in the Minkowski spacetime as
\begin{align} & 
G_W(x, x') = \frac{- 1}{( 2 \pi )^2} 
\frac{1}{( t - t' - \ImUnit \, \varepsilon )^2 - | \evec{x} - \evec{x}' |^2} , 
\label{eqn:WightmanMinScalar} \\ & 
G_F(x , x') = \frac{\ImUnit}{( 2 \pi )^2} 
\frac{1}{(t - t')^2 - | \evec{x}' - \evec{x} |^2 - \ImUnit \varepsilon} , 
\label{eqn:FeymnamProExplicit} 
\end{align} 
with $\varepsilon > 0$. 
We note here that Eqs. (\ref{eqn:CalETotDef}) and (\ref{eqn:FeymnamProExplicit}) explicitly show that $\mathcal{I}_E$ is 
symmetric under the exchange of the roles of Alice and Bob, $A \leftrightarrow B$. 

As in the case of Ref. \cite{{Reznik03}}, we have shown in Ref. \cite{KogaKM18} also for the general monopole coupling, the positive partial transpose (PPT) criterion \cite{peres,HorodeckiHH96} implies that the two detectors are entangled at the asymptotic future, if the condition 
\begin{equation} 
\mathcal{P}_A \, \mathcal{P}_B < \left| \mathcal{E} \right|^2 , 
\label{eqn:EntanglementCond1} 
\end{equation} 
is satisfied. Moreover, 
we have computed the optimal fidelity for the standard quantum teleportation \cite{HorodeckiHH96b}, and have shown that 
the entanglement extracted from the vacuum in this manner is usable in the standard quantum teleportation 
in the symmetric case $\mathcal{P}_A = \mathcal{P}_B$ \cite{KogaKM18}. 
There exists another possibility for entanglement extraction, which does not occur when Eq. (\ref{eqn:EntanglementCond1}) holds, 
but it has been shown that the standard quantum teleportation is not possible in the latter case. 
We thus focus in this paper on the condition Eq. (\ref{eqn:EntanglementCond1}) to investigate 
whether usable entanglement is extracted from the Minkowski vacuum into the pair of Unruh-DeWitt detectors. 
We see that the condition Eq. (\ref{eqn:EntanglementCond1}) 
for entanglement is rephrased, 
in term of the integrals in Eqs. (\ref{eqn:CalIIDef}) and (\ref{eqn:CalETotDef}), as the inequality 
\begin{equation}
\mathcal{I}_A \, \mathcal{I}_B < \left| \mathcal{I}_E \right|^2 . 
\label{eqn:EntanglementCondIneq} 
\end{equation} 
We note here that since $\mathcal{P}_I$ is the excitation probability of the detector $I$, $\mathcal{P}_I \geq 0$, and thus 
\begin{equation}
\mathcal{I}_I \geq 0 . 
\label{eqn:PositivityCalII} 
\end{equation} 
Therefore, we immediately see that usable entanglement is not extracted in the case of $\mathcal{I}_E = 0$. 
As we see from Eq. (\ref{eqn:CalETotDef}), $\mathcal{I}_E$ describes the cross-correlation between Alice and Bob, 
and thus Eq. (\ref{eqn:EntanglementCondIneq}) is interpreted as stating that the detectors are entangled when the correlation larger 
than the excitation probabilities arises between them.

When entanglement is found to be extracted, 
it is then of interest to quantify the amount of the entanglement. 
The negativity $\mathcal{N}(\rho_{AB})$ of a state $\rho_{AB}$, defined as the sum of the negative eigenvalues of the partial transpose of 
$\rho_{AB}$, 
is known to give an upper bound of distillable entanglement \cite{UpBoundDis}. 
For the density matrix (\ref{eqn:RhoAB}), the negativity $\mathcal{N}(\rho_{AB})$ has been derived, when the condition for 
entanglement (\ref{eqn:EntanglementCond1}) is satisfied, as \cite{KogaKM18} 
\begin{equation}
\mathcal{N}(\rho_{AB}) = - \dfrac{c^2}{2} \left[ \mathcal{P}_A + \mathcal{P}_B 
- \sqrt{\left( \mathcal{P}_A - \mathcal{P}_B \right)^2 + 4 \left| \mathcal{E} \right|^2} \right] + O(c^4) . 
\label{eqn:Negativity} 
\end{equation} 
A more operationally meaningful measure of entanglement is the entanglement of formation $E_F(\rho_{AB})$ \cite{BennettDSW96}, the minimum of 
the entanglement entropy when the state $\rho_{AB}$ is decomposed into pure states. Although the entanglement of formation $E_F(\rho_{AB})$ is generally difficult to calculate, it is related with the concurrence $C(\rho_{AB})$ in the case of two qubits, as \cite{Wootters98}  
\begin{equation}
E_F(\rho_{AB}) = h\left( \frac{1 + \sqrt{1-C^2(\rho_{AB})}}{2} \right) , 
\end{equation}
where $h(x) \equiv -x \log x - (1-x) \log(1-x)$ is the binary entropy, and the concurrence $C(\rho_{AB})$ for two-qubit is defined by 
\begin{equation}
C(\rho_{AB}) \equiv \max\left[0, \tilde{\lambda}_1 - \tilde{\lambda}_2 - \tilde{\lambda}_3 - \tilde{\lambda}_4 \right] , 
\label{eqn:ConcurrenceDef} 
\end{equation}
where $\tilde{\lambda}_1$, $\tilde{\lambda}_2$, $\tilde{\lambda}_3$, and $\tilde{\lambda}_4$ are the square roots of the eigenvalues of the matrix 
$\rho_{AB} \, \sigma_y \otimes \sigma_y \, \rho^\ast_{AB} \, \sigma_y \otimes \sigma_y$ in the descending order. 
Under the condition (\ref{eqn:EntanglementCond1}) for entanglement, the concurrence $C(\rho_{AB})$ for the density matrix (\ref{eqn:RhoAB}) 
has been derived as \cite{KogaKM18}
\begin{equation}
C(\rho_{AB}) = 2 \, c^2 \left( |{\cal E}| - \sqrt{{\cal P}_A{\cal P}_B} \right) + O(c^4) 
\label{eqn:Concurrence} 
\end{equation} 
In particular, in the symmetric case $\mathcal{P}_A = \mathcal{P}_B$, where $\mathcal{I}_A = \mathcal{I}_B \equiv \mathcal{I}$ holds 
and we additionally assume 
\begin{equation} 
\bra{E_{1}^{(A)}} m_A(0) \cket{E_0^{(A)}} = \bra{E_{1}^{(B)}} m_B(0) \cket{E_0^{(B)}} 
\equiv \bra{E_{1}} m(0) \cket{E_0} , 
\label{eqn:AssMatrixEleSymmetric} 
\end{equation} 
the negativity in Eq. (\ref{eqn:Negativity}) and 
the concurrence in Eq. (\ref{eqn:Concurrence}) are related as \cite{MartinMartinezST16,KogaKM18}
\begin{equation}
C(\rho_{AB}) = 2 \, {\cal N}(\rho_{AB}) 
= 2 \, c^2 \left| \bra{E_{1}} m(0) \cket{E_0} \right|^2 \left( | \mathcal{I}_E | - \mathcal{I} \right) + O(c^4) .  
\label{eqn:ConcurrenceNegativity} 
\end{equation} 

For uniformly accelerated detectors $I$, as we consider in this paper, $\mathcal{I}_I$ is derived as \cite{DeWitt79,BirrellDavies82}  
\begin{align} & 
\mathcal{I}_I 
= \frac{\Delta E^{(I)}}{2 \pi} \frac{1}{e^{2 \pi \frac{\Delta E^{(I)}}{\kappa_I}} - 1} 
\int_{- \infty}^{\infty} d \tau'_I , 
\label{eqn:CalIIAnitAccel} 
\end{align} 
where $\kappa_I$ is the magnitude of the four-acceleration of the detector $I$. 
Although Eq. (\ref{eqn:CalIIAnitAccel}) implies that the detector is excited in accord with Planckian distribution of the temperature $\kappa_I / 2 \pi$, Eq. (\ref{eqn:CalIIAnitAccel}), and hence the excitation probability $\mathcal{P}_I$, diverge, when taken as literally. 
However, this occurs because we have set the switching function as $\chi_I(\tau_I) = 1$, without specifying the details of adiabatic switching 
at the asymptotic past and future. 
Instead, 
as in the case of Fermi's golden rule,  
one usually considers 
the excitation rate per unit proper time $\dot{\mathcal{P}}_I$ \cite{DeWitt79,BirrellDavies82}. Up to the matrix element of $m_I(0)$, 
we thus consider  
\begin{equation} 
\dot{\mathcal{I}}_I = \frac{\Delta E^{(I)}}{2 \pi} \frac{1}{e^{2 \pi \frac{\Delta E^{(I)}}{\kappa_I}} - 1} . 
\label{eqn:ExciteRate} 
\end{equation} 
Furthermore, the perturbative calculation based on which Eq. (\ref{eqn:CalIIAnitAccel}) is derived will break down, 
before $\mathcal{I}_I$ diverges. 
On the other hand, within the perturbative regime we are interested in, 
which requires taking the formal limit of $c \rightarrow 0$ or adiabatic switching,  
thermal equilibrium of the detector with the thermal bath 
(for a uniformly accelerated observer) has not been achieved yet even at sufficiently late time, and thus the detector 
is kept being excited at the constant excitation rate (\ref{eqn:ExciteRate}). 
We then see, within the perturbative regime, that if $\left| \mathcal{I}_E \right|$ remains smaller than $\mathcal{I}_I$, which grows linearly in the proper time with 
the rate Eq. (\ref{eqn:ExciteRate}), the condition Eq. (\ref{eqn:EntanglementCondIneq}) for entanglement is not satisfied at the asymptotic future. 
Therefore, roughly speaking, in order for the detectors to be entangled, 
the cross-correlation described by  
$\left| \mathcal{I}_E \right|$ needs to formally diverge at least as ``fast'' as the excitation probability $\mathcal{I}_I$, 
whose precise condition will be investigated below in explicit cases. For that purpose, it will be convenient to write Eq. (\ref{eqn:CalIIAnitAccel}) as 
\begin{align} & 
\mathcal{I}_I 
= \frac{1}{2 \pi} \frac{\Delta E^{(I)}}{\kappa_I} \frac{1}{e^{2 \pi \frac{\Delta E^{(I)}}{\kappa_I}} - 1} 
\int_{- \infty}^{\infty} d \lambda_I , 
\label{eqn:CalIIAnitAccelPr} 
\end{align} 
by introducing the dimensionless affine parameter defined as 
\begin{equation} 
\lambda_I = \kappa_I \, \tau_I' . 
\label{eqn:LamndaIDef} 
\end{equation}

\section{Entanglement extraction} 
\label{sec:entangle} 

\subsection{Framework} 

We employ an inertial coordinate system, such that Alice is accelerated in the $x$-direction. 
By denoting the magnitude of the acceleration of Alice as $\kappa_A$, the coordinates of the worldline of 
Alice are written as  
\begin{equation} 
\bar{t}_A(\tau_A) = \frac{1}{\kappa_A} \, \sinh \left( \kappa_A \tau_A \right) , \qquad 
\bar{x}_A(\tau_A) = \frac{1}{\kappa_A} \, \cosh \left( \kappa_A \tau_A \right) , \qquad 
\bar{y}_A(\tau_A) = 0 , \qquad \bar{z}_A(\tau_A) = 0 .  
\label{eqn:WorldLineAlice} 
\end{equation} 
Since the worldline coordinates above depend on the proper time $\tau_A$ of Alice only through the hyperbolic functions, 
the spacetime interval between Alice at $\tau_A$ and Bob at $\tau_B$ (the denominator of $G_F(\bar{x}_B , \bar{x}_A)$ in 
Eq. (\ref{eqn:FeymnamProExplicit}) 
with $\varepsilon = 0$) is generally written as 
\begin{equation} 
\left( \bar{t}_B(\tau_B) - \bar{t}_A(\tau_A) \right)^2 - \left| \bar{\boldsymbol{x}}_B(\tau_B) - \bar{\boldsymbol{x}}_A(\tau_A) \right|^2  
=  K \left[ A(\tau_B) \, e^{\kappa_A \tau_A} - 2 B(\tau_B) + C(\tau_B) \, e^{- \kappa_A \tau_A}  \right] , 
\label{eqn:CausalityGen} 
\end{equation} 
where $A(\tau_B)$, $B(\tau_B)$, and $C(\tau_B)$ are functions of Bob's proper time $\tau_B$, and $K$ is a constant.  
Then, the denominator of $G_F(\bar{x}_B , \bar{x}_A)$ is written as 
\begin{align} & 
\left( \bar{t}_B(\tau_B) - \bar{t}_A(\tau_A) \right)^2 - \left| \bar{\boldsymbol{x}}_B(\tau_B) - \bar{\boldsymbol{x}}_A(\tau_A) \right|^2 - \ImUnit \varepsilon 
= K A(\tau_B) \, e^{- \kappa_A \tau_A} \left( e^{\kappa_A \tau_A} - \eta_+(\tau_B) \right) 
\left( e^{\kappa_A \tau_A} - \eta_-(\tau_B) \right) , 
\label{eqn:DenomGFAntiAccelLg} 
\end{align} 
where 
\begin{align} & 
\eta_{\pm}(\tau_B) = \frac{1}{A(\tau_B)} \left[ B(\tau_B) \pm D(\tau_B) \right] \left[ 1 
\pm \ImUnit \, \varepsilon \, \frac{1}{2 K} \frac{1}{D(\tau_B)} 
\right] , 
\label{eqn:EtaPMGen} 
\end{align} 
and 
\begin{equation}
D(\tau_B) \equiv \sqrt{B^2(\tau_B) - A(\tau_B) \, C(\tau_B)} \quad \mathrm{or} \quad - \sqrt{B^2(\tau_B) - A(\tau_B) \, C(\tau_B)}  . 
\label{eqn:DDefGen} 
\end{equation}
Note that one can choose either sign in Eq. (\ref{eqn:DDefGen}), because of the symmetry under the interchange $\eta_+ \leftrightarrow \eta_-$.  

We then substitute Eqs. (\ref{eqn:FeymnamProExplicit}) and (\ref{eqn:DenomGFAntiAccelLg}) into Eq. (\ref{eqn:CalETotDef}), 
and perform the integration by $\tau_A$, by considering the infinite semi-circle in the upper half of the complex $\tau_A$ plane, as 
\begin{equation}
\mathcal{I}_E = \frac{\ImUnit}{4 \pi K} \frac{1}{\kappa_A} \:  \int_{- \infty}^{\infty} d \tau_B \, 
\frac{e^{\ImUnit \, \Delta E^{(B)} \tau_B}}{D(\tau_B)} 
\left\{ \sum_n e^{\ImUnit \Delta E^{(A)} \tau_{A+n}(\tau_B)} - \sum_n e^{\ImUnit \Delta E^{(A)} \tau_{A-n}(\tau_B)} \right\} , 
\label{eqn:CalIEFirstIntGen} 
\end{equation}
where the poles of the integrand are located at 
\begin{equation}
\tau_A = \tau_{A\pm n}(\tau_B) \equiv \frac{1}{\kappa_A} \ln \eta_{\pm}(\tau_B) + \frac{1}{\kappa_A} 2 \pi n \ImUnit , 
\label{eqn:TauAPMNDefGen} 
\end{equation} 
and the summation over $n$ in Eq. (\ref{eqn:CalIEFirstIntGen}) is restricted such that these poles are encircled by 
the infinite semi-circle in the upper half of the complex $\tau_A$ plane, and thus the imaginary parts of $\tau_{A\pm n}(\tau_B)$ are restricted to be positive. 
By deforming the integral contour near the poles without crossing the poles, the imaginary part in the square bracket in Eq. (\ref{eqn:EtaPMGen}) 
can be set to be a constant, as long as its sign is kept unchanged. 
In what follows, we will consider explicit examples where one can investigate entanglement extraction analytically, based on 
Eqs. (\ref{eqn:EntanglementCondIneq}), (\ref{eqn:CalIIAnitAccelPr}) and (\ref{eqn:CalIEFirstIntGen}).

\subsection{Parallel acceleration} 

We first consider the case where Bob is uniformly accelerated in the same direction as Alice, i.e., parallelly accelerated to Alice.  
If Bob follows the same worldline as Alice, it is natural to expect that the two detectors will be entangled. 
Thus, we will assume the two worldlines are translationally shifted. We will also consider the case where the two detectors are uniformly accelerated 
with the different magnitudes of acceleration. 

\subsubsection{Transverse shift} 

We begin with the case where Alice and Bob are accelerated parallelly with the same magnitude $\kappa$ of the acceleration, 
with a translational shift in the transverse direction with respect to the acceleration. The worldline of Bob is then described by  
\begin{equation} 
\bar{t}_B(\tau_B) = \frac{1}{\kappa} \, \sinh \left( \kappa \tau_B \right) , \qquad 
\bar{x}_B(\tau_B) = \frac{1}{\kappa} \, \cosh \left( \kappa \tau_B \right) , \qquad 
\bar{y}_B(\tau_B) = y_0 , \qquad \bar{z}_B(\tau_B) = z_0 , 
\label{eqn:WorldlineBParaAccel} 
\end{equation} 
where $y_0$ and $z_0$ are arbitrary non-vanishing constants, while Alice's worldline is given by Eq. (\ref{eqn:WorldLineAlice}) with $\kappa_A = \kappa$. 
\begin{figure} 
\includegraphics[scale=0.5,keepaspectratio]{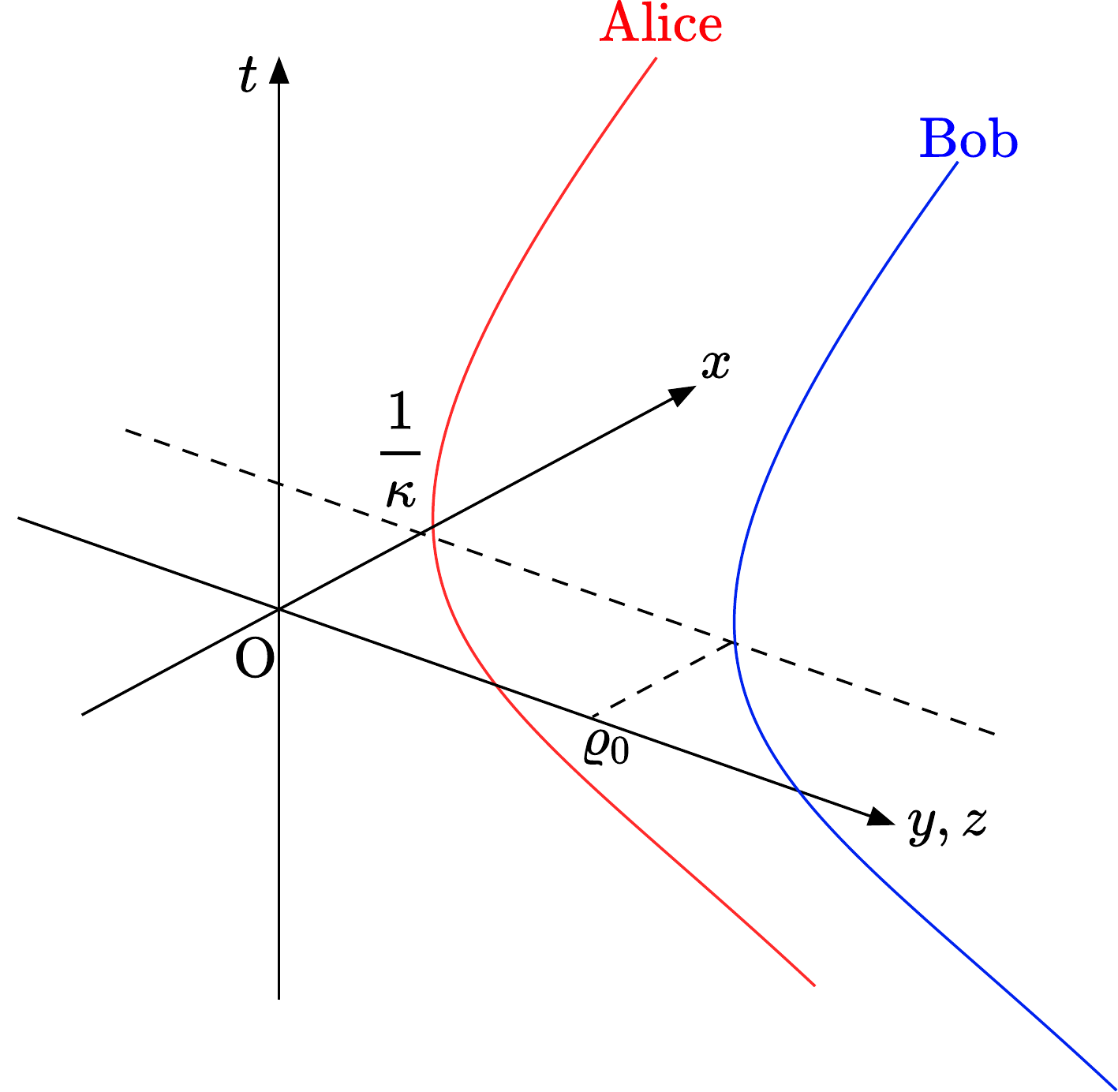} 
\caption{The worldlines of Alice and Bob accelerated parallelly with a translational shift transverse to the acceleration.} 
\end{figure} 
The spacetime interval in this case is given by Eq. (\ref{eqn:CausalityGen}) with $A(\tau_B)$, $B(\tau_B)$, $C(\tau_B)$, and $K$ being given by 
\begin{align} & 
A(\tau_B) \equiv e^{- \kappa \tau_B}  , \qquad 
B(\tau_B) \equiv 1 + \frac{\kappa^2 \varrho_0^2}{2} , \qquad 
C(\tau_B) \equiv e^{\kappa \tau_B}   , \qquad 
K \equiv \frac{1}{\kappa^2} . 
\label{eqn:KDefParaAccel} 
\end{align} 
where $\varrho_0$ is the constant larger than zero defined by  
\begin{equation}
\varrho_0 \equiv \sqrt{y_0^2 + z_0^2}  . 
\label{eqn:Rho0Def} 
\end{equation} 
We then derive $D(\tau_B)$ defined in Eq. (\ref{eqn:DDefGen}) as 
\begin{equation} 
D(\tau_B) 
= \kappa \varrho_0 \sqrt{1 + \frac{\kappa^2 \varrho_0^2}{4}}  > 0 , 
\label{eqn:DDefParaAccel} 
\end{equation} 
and $\tau_{A\pm n}$ defined in Eq. (\ref{eqn:TauAPMNDefGen}) as 
\begin{equation} 
\tau_{A\pm n}(\tau_B)  
= \tau_B + \frac{1}{\kappa} \ln \varpi_{\pm} + \frac{1}{\kappa} 2 \pi n \ImUnit \pm \ImUnit \, \varepsilon , 
\label{eqn:TauAPMNParaAccel} 
\end{equation} 
where 
\begin{equation}
\varpi_{\pm} \equiv 1 + \frac{\kappa^2 \varrho_0^2}{2} \pm \kappa \varrho_0 \sqrt{1 + \frac{\kappa^2 \varrho_0^2}{4}} .  
\end{equation}
By substituting (\ref{eqn:KDefParaAccel}), (\ref{eqn:DDefParaAccel}), and (\ref{eqn:TauAPMNParaAccel}) into Eq. (\ref{eqn:CalIEFirstIntGen}), 
we compute $\mathcal{I}_E$ as 
\begin{equation} 
\mathcal{I}_E = \frac{\ImUnit \kappa}{\varpi_+ - \varpi_-} \frac{1}{1 - e^{- 2 \pi \frac{\Delta E^{(A)}}{\kappa}}} \: 
\left[ \varpi_+^{\ImUnit \frac{\Delta E^{(A)}}{\kappa}} 
- e^{- 2 \pi \frac{\Delta E^{(A)}}{\kappa}} \varpi_-^{\ImUnit \frac{\Delta E^{(A)}}{\kappa}} \right] \: 
\delta \left( \Delta E^{(A)} + \Delta E^{(B)} \right) = 0 ,  
\label{eqn:CalIEParaAccel} 
\end{equation} 
where one notes Eq. (\ref{eqn:NonDegenerate}) in the last equality. 
Thus, we see from Eqs. (\ref{eqn:EntanglementCondIneq}), (\ref{eqn:PositivityCalII}), and (\ref{eqn:CalIEParaAccel}) that the two detectors are not entangled in this case. As in the case of a comoving inertial motion \cite{KogaKM18}, the appearance of the delta function in Eq. (\ref{eqn:CalIEParaAccel}) 
is understood as the indication of the energy conservation. 
It is interesting that the energy conservation plays a decisive role even in this case where energy is imparted into the system 
in order to accelerate the two detectors.

\subsubsection{Different acceleration} 

As the next example of parallelly accelerated detectors, we here consider the two detectors accelerated with different magnitudes of 
the acceleration. The worldline coordinates of Alice are thus given by Eq. (\ref{eqn:WorldLineAlice}), and Bob's worldline is assumed to be 
described by 
\begin{equation}
\bar{t}_B(\tau_B) = \frac{1}{\kappa_B} \, \sinh \left( \kappa_B \tau_B \right) , \qquad 
\bar{x}_B(\tau_B) = \frac{1}{\kappa_B} \, \cosh \left( \kappa_B \tau_B \right)  , \qquad 
\bar{y}_B(\tau_B) = 0 , \qquad \bar{z}_B(\tau_B) = 0 , 
\label{eqn:WorldlinesParaAccelDif} 
\end{equation}
where we assume $\kappa_A > \kappa_B$, without loss of generality. 
\begin{figure} 
\includegraphics[scale=0.5,keepaspectratio]{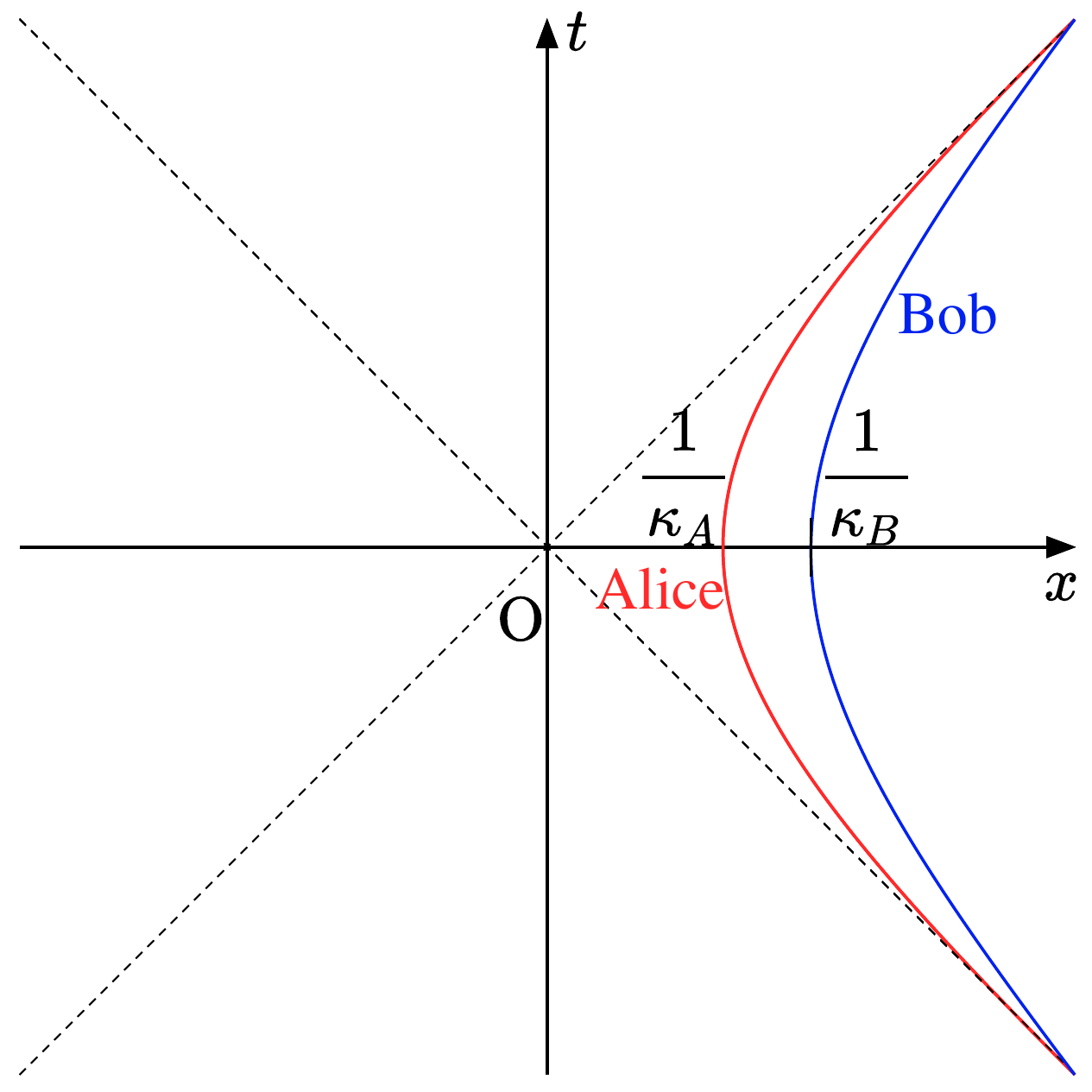} 
\caption{The worldlines of Alice and Bob accelerated parallelly with different magnitude of the acceleration.} 
\end{figure} 
The spacetime interval Eq. (\ref{eqn:CausalityGen}) in this case is described by 
\begin{equation} 
A(\tau_B) \equiv e^{- \kappa_B \tau_B}  , \qquad 
B(\tau_B) \equiv \frac{1}{2} \left( \frac{\kappa_A}{\kappa_B} + \frac{\kappa_B}{\kappa_A} \right) 
= \cosh \sigma , \qquad 
C(\tau_B) \equiv e^{\kappa_B \tau_B}  , \qquad 
K \equiv \frac{1}{\kappa_A \kappa_B} ,
\label{eqn:KDefParaAccelDif} 
\end{equation} 
where 
\begin{equation}
\sigma \equiv \ln \frac{\kappa_A}{\kappa_B} > 0 . 
\label{eqn:SigmaDefParaAccelDif} 
\end{equation} 
We then obtain 
\begin{equation} 
D(\tau_B) 
= \frac{1}{2} \left( \frac{\kappa_A}{\kappa_B} - \frac{\kappa_B}{\kappa_A} \right) 
= \sinh \sigma , 
\label{eqn:DDefParaAccelDif}  
\end{equation} 
and 
\begin{equation}
\tau_{A\pm n}(\tau_B) 
= \frac{\kappa_B}{\kappa_A} \tau_B \pm \frac{\sigma}{\kappa_A} + \frac{1}{\kappa_A} 2 \pi n \ImUnit \pm \ImUnit \, \varepsilon . 
\label{eqn:TauAPMNDefParaAccelDif} 
\end{equation} 
From Eqs. (\ref{eqn:CalIEFirstIntGen}), (\ref{eqn:KDefParaAccelDif}), (\ref{eqn:DDefParaAccelDif}), and (\ref{eqn:TauAPMNDefParaAccelDif}), 
$\mathcal{I}_E$ is derived as 
\begin{align} & 
\mathcal{I}_E 
= \frac{\ImUnit}{2 \sinh \sigma} \frac{1}{1 - e^{- 2 \pi \frac{\Delta E^{(A)}}{\kappa_A}}} 
\left\{ e^{\ImUnit \frac{\Delta E^{(A)}}{\kappa_A} \sigma} 
- e^{- \ImUnit \frac{\Delta E^{(A)}}{\kappa_A} \sigma} 
e^{- 2 \pi \frac{\Delta E^{(A)}}{\kappa_A}} \right\} \:  \delta \left( \frac{\Delta E^{(B)}}{\kappa_B} + \frac{\Delta E^{(A)}}{\kappa_A} \right) = 0 . 
\label{eqn:CalIEFirstIntParaAccelDif} 
\end{align} 
We immediately see from Eqs. (\ref{eqn:EntanglementCondIneq}), (\ref{eqn:PositivityCalII}), and (\ref{eqn:CalIEFirstIntParaAccelDif}) that no entanglement is extracted from the vacuum. 
We also see that the delta function appears again in Eq. (\ref{eqn:CalIEFirstIntParaAccelDif}), but now its argument is the sum of 
the ratio $\Delta E^{(I)} / \kappa_I$ of the excitation energy to the magnitude of the acceleration, not the excitation energy itself. 
However, this is still understood as the energy conservation. 
To see this, we consider the timelike boost Killing vector along the worldline of the detector $I$,  
which is written, by using Eqs. (\ref{eqn:WorldLineAlice}) and (\ref{eqn:WorldlinesParaAccelDif}),  as 
\begin{equation} 
\left. x \, \frac{\partial}{\partial t} + t \, \frac{\partial}{\partial x} \right|_{x^{\mu} = \bar{x}_I^{\mu}(\tau_I)} 
= \frac{1}{\kappa_I} \frac{d}{d \tau_I} .  
\label{eqn:BoostKilling} 
\end{equation} 
This Killing vector is nothing but the timelike Killing vector in the Rindler chart, which 
defines the energy of the Rindler mode.
The facts that the excitation energy $\Delta E^{(I)}$ is defined with respect to the proper time $\tau_I$, and that 
the energy associated with a timelike Killing vector is conserved explain the appearance of the ratio 
$\Delta E^{(I)} / \kappa_I$ in the delta function. 
Therefore, we see again that the energy conservation forbids entanglement extraction.

\subsubsection{Longitudinal Shift} 

Thirdly, we suppose that Bob follows a parallelly accelerated worldline that is shifted longitudinally, i.e., in the same spatial direction as the acceleration. 
The magnitudes of the accelerations of two detectors are assumed to be the same, which we denote by $\kappa$.  
Bob's worldline is thus given as 
\begin{equation}
\bar{t}_B(\tau_B) = \frac{1}{\kappa} \, \sinh \left( \kappa \tau_B \right) , \qquad 
\bar{x}_B(\tau_B) = \frac{1}{\kappa} \, \cosh \left( \kappa \tau_B \right) + x_0 , \qquad 
\bar{y}_B(\tau_B) = 0 , \qquad \bar{z}_B(\tau_B) = 0 , 
\label{eqn:WorldlinesParaAccelLgShift} 
\end{equation}
where we assume the constant $x_0$ is larger than zero, without loss of generality.  
The worldline coordinates of Alice are given by Eq. (\ref{eqn:WorldLineAlice}) with $\kappa_A = \kappa$, again. 
\begin{figure} 
\includegraphics[scale=0.5,keepaspectratio]{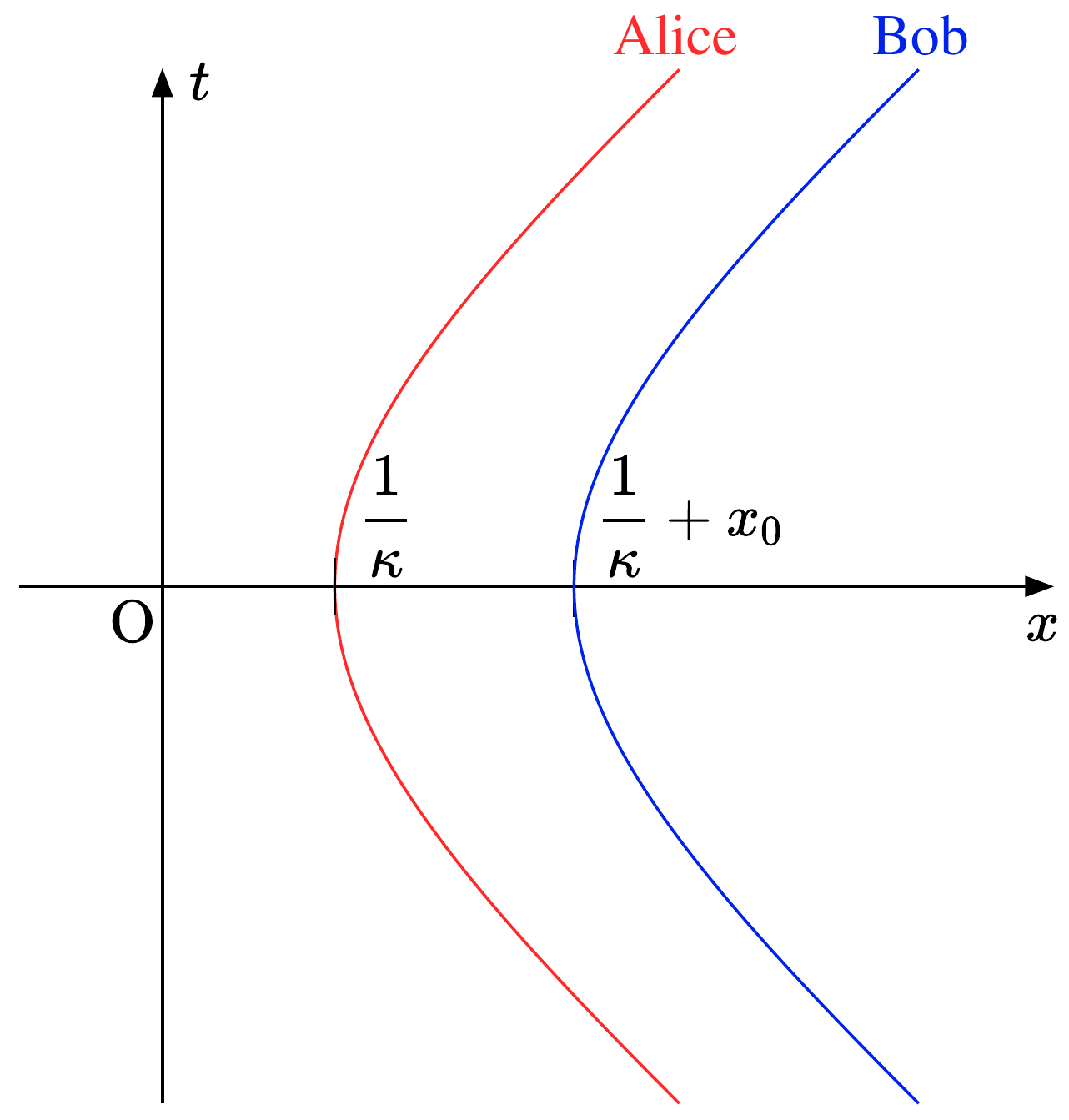} 
\caption{The worldlines of Alice and Bob accelerated parallelly with a longitudinal shift to the acceleration.} 
\end{figure} 
In this case, the functions $A(\tau_B)$, $B(\tau_B)$, and $C(\tau_B)$, and the constant $K$ in 
the spacetime interval (\ref{eqn:CausalityGen}) are derived as 
\begin{equation}
A(\tau_B) \equiv e^{- \kappa \tau_B} + \kappa x_0  , \qquad 
B(\tau_B) \equiv \kappa x_0  \cosh \left( \kappa \tau_B \right) + \frac{\kappa^2 x_0^2}{2} +1 , \qquad 
C(\tau_B) \equiv e^{\kappa \tau_B} + \kappa x_0  , \qquad 
K \equiv \frac{1}{\kappa^2} ,  
\label{eqn:KDefParaAccelLg} 
\end{equation}
from which we obtain 
\begin{equation}
D(\tau_B) 
= \kappa x_0 \cosh \left( \kappa \tau_B \right) + \frac{\kappa^2 x_0^2}{2} > 0 , 
\label{eqn:DDefParaAccelLg} 
\end{equation}
and 
\begin{equation} 
\tau_{A\pm n}(\tau_B) 
= \pm \frac{1}{\kappa} \ln \left( e^{\pm \kappa \tau_B} + \kappa x_0 \right) + \frac{1}{\kappa} 2 \pi n \ImUnit \pm \ImUnit \, \varepsilon .  
\label{eqn:TauAPMNDefParaAccelLg} 
\end{equation} 
From Eqs. (\ref{eqn:CalIEFirstIntGen}), (\ref{eqn:KDefParaAccelLg}), (\ref{eqn:DDefParaAccelLg}), and (\ref{eqn:TauAPMNDefParaAccelLg}),  
we compute as 
\begin{align} & 
\mathcal{I}_E 
= \frac{\ImUnit}{2 \pi x_0} \frac{1}{1 - e^{- 2 \pi \frac{\Delta E^{(A)}}{\kappa}}} \:  \int_{- \infty}^{\infty} d \tau_B \, 
\frac{e^{\ImUnit \, \Delta E^{(B)} \tau_B}}{2 \cosh \left( \kappa \tau_B \right) + \kappa x_0} 
\left[ e^{\ImUnit \frac{\Delta E^{(A)}}{\kappa} \ln \left( e^{\kappa \tau_B} + \kappa x_0 \right)}  
- e^{- 2 \pi \frac{\Delta E^{(A)}}{\kappa}} \, e^{- \ImUnit \frac{\Delta E^{(A)}}{\kappa} \ln \left( e^{- \kappa \tau_B} + \kappa x_0 \right)} 
\right]  . 
\label{eqn:CalIEFirstIntParaAccelLg} 
\end{align} 
Then the triangle inequality and the integral inequality yield 
\begin{equation}
\left| \mathcal{I}_E \right| \leq 
\frac{1}{2 \pi \kappa x_0} \coth \left( \pi \frac{\Delta E^{(A)}}{\kappa} \right) P , 
\label{eqn:CalIEParaAccelBnd} 
\end{equation}
where 
\begin{align} & 
P \equiv \int_{- \infty}^{\infty} d \tau_B \, \frac{\kappa}{2 \cosh \left( \kappa \tau_B \right) + \kappa x_0} 
= \left\{ \begin{array}{lll} 
\dfrac{1}{\sqrt{\dfrac{\kappa^2 x_0^2}{4} - 1}} \; \ln \left( \dfrac{\kappa x_0}{2} + \sqrt{\dfrac{\kappa^2 x_0^2}{4} - 1} \right) 
& \quad \mathrm{for} \quad & \dfrac{\kappa x_0}{2} > 1 \\ \\ 
1 & \quad \mathrm{for}  \quad & \dfrac{\kappa x_0}{2} = 1 \\ 
\dfrac{1}{\sqrt{1 - \dfrac{\kappa^2 x_0^2}{4}}} \; \arctan \dfrac{\sqrt{1 - \dfrac{\kappa^2 x_0^2}{4}}}{\dfrac{\kappa x_0}{2}}
& \quad \mathrm{for}  \quad & \dfrac{\kappa x_0}{2} < 1
\end{array} \right. . 
\end{align} 
Since Eq. (\ref{eqn:CalIEParaAccelBnd}) shows that $\mathcal{I}_E$ is bounded and hence remains smaller than $\mathcal{I}_I$, 
which grows linearly in the proper time, at a sufficiently late time, 
we see that the two detectors are not entangled in this case, either. 
However, $\mathcal{I}_E$ may not vanish in this case, in contrast to the above two cases. 
The delta function does not appear in Eq. (\ref{eqn:CalIEParaAccelBnd}). 
This might seem odd, since the two detectors look ``comoving'' in this case. 
However, the Rindler charts and hence the Rindler modes associated with Alice and Bob respectively are different between each other in this case, actually. This corresponds to the fact that the factor $x$ in the first term of the left-hand side of the boost Killing vector in Eq. (\ref{eqn:BoostKilling}) should be shifted by $x_0$ for Bob.Therefore, the timelike Killing vector fields that define the energies are different between Alice and Bob, and thus the energy conservation is not simple enough to be expressed in terms of the delta function in $\mathcal{I}_E$.

\subsection{Anti-parallel acceleration}  

Now we turn to the case where Bob is uniformly accelerated in the opposite direction to Alice, in other words, 
the case where Alice and Bob are accelerated anti-parallelly. 
We will first exhibit the case where entanglement extraction is possible. 
However, it is not always the case. We will also show that entanglement is not extracted if the worldlines are translationally shifted 
longitudinally to the spatial direction of the acceleration. 

\subsubsection{Entangled case} 
\label{sec:Entangled} 

We here consider that Bob is accelerated anti-parallelly to Alice with an arbitrary magnitude 
of the acceleration, and with a possible transverse shift, where the worldline coordinates of Bob are given by 
\begin{equation}
\bar{t}_B(\tau_B) = \frac{1}{\kappa_B} \sinh \left( \kappa_B \tau_B \right) , \quad 
\bar{x}_B(\tau_B) = - \frac{1}{\kappa_B} \cosh \left( \kappa_B \tau_B \right)  , \quad 
\bar{y}_B(\tau_B) = y_0 , \quad \bar{z}_B(\tau_B) = z_0 , 
\label{eqn:WorldLinesAntiAccelTrShift} 
\end{equation}
with $y_0$ and $z_0$ being arbitrary constants, while the worldline of Alice is described by Eq. (\ref{eqn:WorldLineAlice}).  
\begin{figure} 
\includegraphics[scale=0.5,keepaspectratio]{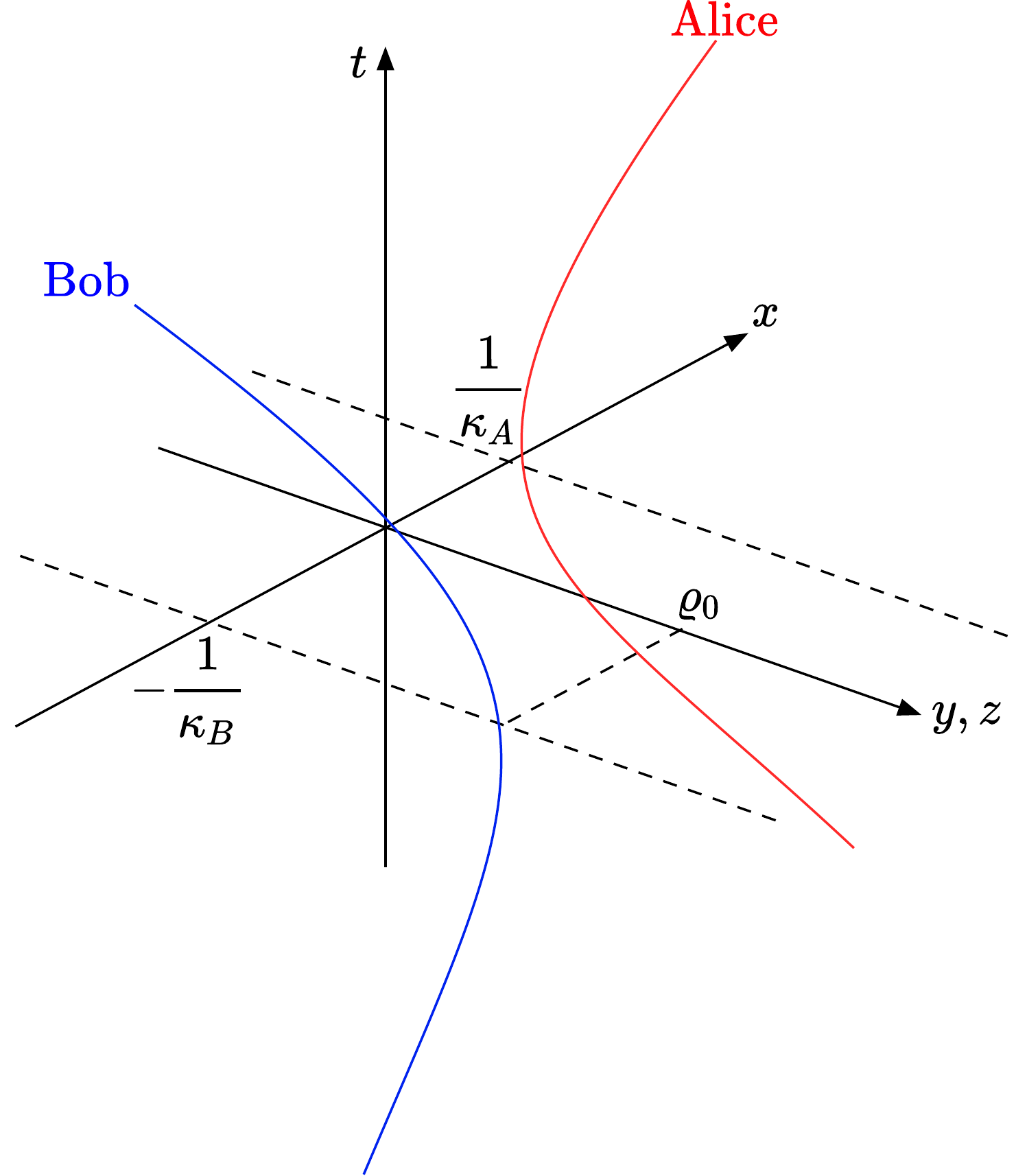} 
\caption{The worldlines of Alice and Bob accelerated anti-parallelly with different magnitudes of acceleration and a transverse shift.} 
\end{figure} 
The functions $A(\tau_B)$, $B(\tau_B)$, and $C(\tau_B)$, and the constant $K$ in the spacetime interval Eq. (\ref{eqn:CausalityGen}) are 
derived in this case as 
\begin{equation} 
A(\tau_B) \equiv e^{\kappa_B \tau_B} , \qquad 
B(\tau_B) \equiv - \frac{1}{2} \left[ \frac{\kappa_A}{\kappa_B} + \frac{\kappa_B}{\kappa_A} + \kappa_A \kappa_B \varrho_0^2 \right]  , \qquad 
C(\tau_B) \equiv e^{- \kappa_B \tau_B} , \qquad 
K \equiv \frac{- 1}{\kappa_A \kappa_B} , 
\label{eqn:KDefAntiAccelTr} 
\end{equation} 
where $\rho_0$ is the same as that defined in Eq. (\ref{eqn:Rho0Def}), but it may vanish in the present case. 
We obtain 
\begin{equation}
D(\tau_B) 
= \sinh \sigma , 
\label{eqn:DVarpiAntiAccelTr} 
\end{equation}
and 
\begin{equation} 
\tau_{A\pm n}(\tau_B)  
= - \frac{\kappa_B}{\kappa_A} \,  \tau_B \mp \frac{\sigma}{\kappa_A}
+ \frac{1}{\kappa_A} ( 2 n + 1) \pi \ImUnit \mp  \ImUnit \, \varepsilon , 
\label{eqn:TauAPMNAntiAccelTr} 
\end{equation} 
where $\sigma$ introduced in Eq. (\ref{eqn:SigmaDefParaAccelDif}) is now generalized as 
\begin{equation}
\sigma \equiv \ln \left[ 
\frac{\frac{\kappa_A}{\kappa_B} + \frac{\kappa_B}{\kappa_A} + \kappa_A \kappa_B \varrho_0^2}{2} 
+ \sqrt{\left( \frac{\frac{\kappa_A}{\kappa_B} + \frac{\kappa_B}{\kappa_A} + \kappa_A \kappa_B \varrho_0^2}{2} \right)^2 - 1} \right] . 
\label{eqn:SigmaDefAntiAccelTr} 
\end{equation} 
Here we see $\sigma \ge 0$ from $\kappa_A / \kappa_B + \kappa_B / \kappa_A \geq 2$. 
From Eqs. (\ref{eqn:CalIEFirstIntGen}), (\ref{eqn:KDefAntiAccelTr}), (\ref{eqn:DVarpiAntiAccelTr}), and (\ref{eqn:TauAPMNAntiAccelTr}), 
we calculate, by using the dimensionless affine parameter $\lambda_B$ defined by Eq. (\ref{eqn:LamndaIDef}), as 
\begin{align} & 
\mathcal{I}_E 
= \frac{- 1}{4 \pi} \:  \frac{\sin \left( \frac{\Delta E^{(A)}}{\kappa_A} \sigma \right)}{\sinh \sigma} 
\frac{1}{\sinh \left( \pi \frac{\Delta E^{(A)}}{\kappa_A} \right)} 
\int_{- \infty}^{\infty} d \lambda_B \; 
e^{\ImUnit \left( \frac{\Delta E^{(B)}}{\kappa_B} - \frac{\Delta E^{(A)}}{\kappa_A} \right) \lambda_B} 
\label{eqn:CalIEResAntiAccelTrPre} \\ & 
= \frac{- 1}{2} \:  \frac{\sin \left( \frac{\Delta E^{(A)}}{\kappa_A} \sigma \right)}{\sinh \sigma} 
\frac{1}{\sinh \left( \pi \frac{\Delta E^{(A)}}{\kappa_A} \right)} 
\; \delta \left( \frac{\Delta E^{(A)}}{\kappa_A} - \frac{\Delta E^{(B)}}{\kappa_B} \right) . 
\label{eqn:CalIEResAntiAccelTr} 
\end{align} 
We see that the delta function appears, again, but its argument is now the {\it difference} of the ratios $\Delta E^{(I)} / \kappa_I$. 
This corresponds to the fact that the anti-parallelly accelerated worldline of Bob is obtained by formally reversing the sign of 
the magnitude of the acceleration for the parallelly accelerated worldline of Bob. 
(By considering the case of $\varrho_0 = 0$ in order to compare with the above cases, 
the worldline of Bob (\ref{eqn:WorldLinesAntiAccelTrShift}) results from reversing the sign of $\kappa_B$ in   
the worldline (\ref{eqn:WorldlinesParaAccelDif}) for the case of parallel acceleration with the different magnitudes.)   
Correspondingly, we see that $\kappa_B$ in the right-hand side of the boost Killing vector (\ref{eqn:BoostKilling}) for Bob is replaced by $- \kappa_B$, 
by substituting Eq. (\ref{eqn:WorldLinesAntiAccelTrShift}) into the left-hand side of Eq. (\ref{eqn:BoostKilling}), 
which results in 
the minus sign in front of $\Delta E^{(B)} / \kappa_B$ in the argument of the delta function in Eq. (\ref{eqn:CalIEResAntiAccelTr}).  
Therefore, also in this case, the appearance of the delta function is understood as arising from the energy conservation. 
We emphasize here that this form of the delta function results from the behavior of the Rindler modes, as we stated above. 
Thus, a pair of Unruh-DeWitt detectors is found to probe also the correlation in the Minkowski vacuum associated with Unruh effect, 
not only the thermal feature probed by a single Unruh-DeWitt detector. 

\begin{figure} 
\includegraphics[scale=0.4,keepaspectratio]{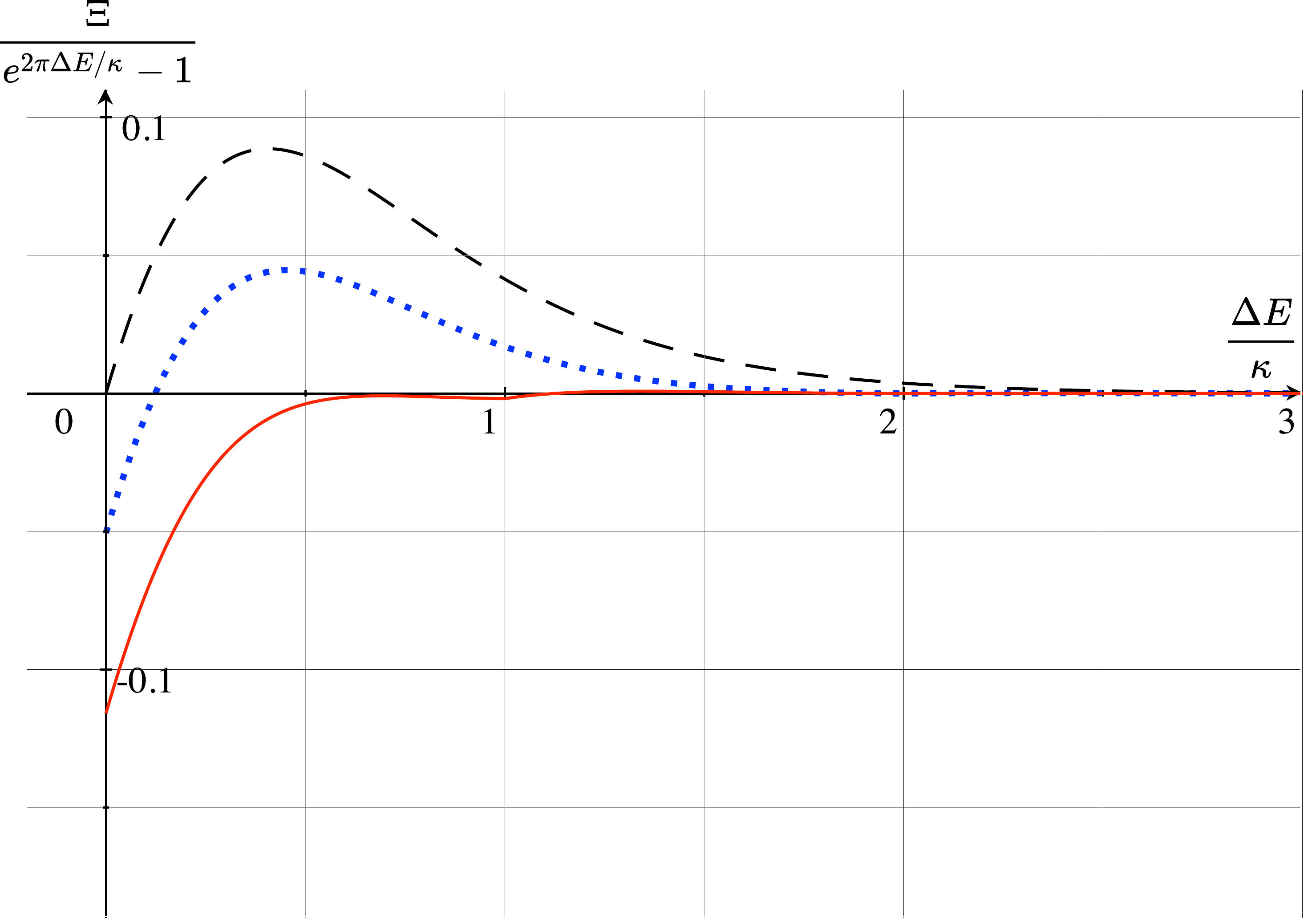} 
\caption{The behavior of $\Xi / \left( \exp \left[ 2 \pi \Delta E / \kappa \right] - 1 \right)$. The black dashed line corresponds to $\sigma = 0$, 
the blue dotted line to $\sigma = \pi / 2$, and the red solid line to $\sigma = \pi$. For sufficiently large value of $\Delta E / \kappa$, 
$\Xi / \left( \exp \left[ 2 \pi \Delta E / \kappa \right] - 1 \right)$ is positive almost everywhere, but extremely small.} 
\label{fig:EntAnti}
\end{figure}

In fact, this leads to the result in sharp contrast with the case of parallel acceleration. 
We see from Eq. (\ref{eqn:CalIEResAntiAccelTr}) that $\mathcal{I}_E$ vanishes when 
$\Delta E^{(A)} / \kappa_A \neq \Delta E^{(B)} / \kappa_B$, and thus entanglement is not extracted in this case. 
However, when $\Delta E^{(A)} / \kappa_A = \Delta E^{(B)} / \kappa_B$, which we write as $\Delta E / \kappa$, 
we find, by using Eqs. (\ref{eqn:CalIIAnitAccelPr}) and (\ref{eqn:CalIEResAntiAccelTrPre}) and recalling that $\mathcal{I}_E$ is symmetric under the exchange between $A$ and $B$,  that the condition (\ref{eqn:EntanglementCondIneq}) for entanglement reduces to 
\begin{equation}
\Xi \equiv \frac{\left| \sin \left( \frac{\Delta E}{\kappa} \sigma \right) \right|}{\sinh \sigma} e^{\pi \frac{\Delta E}{\kappa}} 
- \dfrac{\Delta E}{\kappa} > 0 .  
\label{eqn:XiDefCondEntAntiAccel} 
\end{equation} 
In particular, when $\sigma = 0$, which is found from Eq. (\ref{eqn:SigmaDefAntiAccelTr}) to occur if and only if  
\begin{equation}
\kappa_A = \kappa_B , \quad \mathrm{and} \quad \varrho_0 = 0 , 
\end{equation} 
i.e., when Bob follows the worldline exactly antipodal to Alice's worldline, 
we have $\Xi > 0$ for any value of $\Delta E / \kappa$, 
and thus the two detectors are entangled. 
Even in the case of $\sigma \neq 0$, entanglement extraction is possible. 
In order to see this, we only need to consider the case of $\sigma > 0$, since $\sigma \geq 0$.  
For $\sigma$ large enough, the first term in 
Eq. (\ref{eqn:XiDefCondEntAntiAccel}) may be smaller than the second term 
if $\Delta E / \kappa$ is sufficiently small, and thus the two detectors are not entangled. 
However, as $\Delta E / \kappa$ increases, the first term dominates the second term, unless the sinusoidal function takes the extremely small value. Therefore, even when $\sigma > 0$, the two detectors are able to be entangled 
if the excitation energy $\Delta E$ is much larger than the magnitude $\kappa$ of the acceleration. 
However, the amount of entanglement is not large. To see this, we consider simply the symmetric case $\mathcal{P}_A = \mathcal{P}_B$, 
by assuming Eq. (\ref{eqn:AssMatrixEleSymmetric}) and $\Delta E^{(A)} = \Delta E^{(B)} \equiv \Delta E$, 
which implies $\mathcal{I}_A = \mathcal{I}_B = \mathcal{I}$, and $\kappa_A = \kappa_B$. 
From Eqs. (\ref{eqn:ConcurrenceNegativity}), (\ref{eqn:CalIIAnitAccelPr}) and (\ref{eqn:CalIEResAntiAccelTrPre}), the concurrence $C(\rho_{AB})$ and the negativity ${\cal N}(\rho_{AB})$ 
are then calculated as 
\begin{align} & 
C(\rho_{AB}) = 2 \, {\cal N}(\rho_{AB}) 
= \max \left[ 0 , \frac{c^2}{\pi}  \left| \bra{E_{1}} m(0) \cket{E_0} \right|^2 
\frac{\Xi}{e^{2 \pi \frac{\Delta E}{\kappa}} - 1} 
\int_{- \infty}^{\infty} d \lambda + O(c^4) \right] . 
\end{align} 
As in the above argument that leads to Eq. (\ref{eqn:ExciteRate}), it is meaningful to consider here 
the entanglement extraction rate per unit proper time \footnote{The proper time here refers to that of either Alice or Bob, which gives the same result because they are symmetric.}, 
given as 
\begin{equation}
\dot{C}(\rho_{AB}) = 2 \dot{{\cal N}}(\rho_{AB}) =  \max \left[ 0 , \frac{c^2 \, \kappa}{\pi}  \left| \bra{E_{1}} m(0) \cket{E_0} \right|^2 
\frac{\Xi}{e^{2 \pi \frac{\Delta E}{\kappa}} - 1}  + O(c^4) \right] . 
\end{equation} 
The behavior of $\Xi / \left( \exp \left[ 2 \pi \Delta E / \kappa \right] - 1 \right)$ is depicted in Fig. \ref{fig:EntAnti}. 
We see that the ``Planck factor'' $\exp \left[ 2 \pi \Delta E / \kappa \right] - 1$ suppresses the entanglement extraction at 
large values of $\Delta E / \kappa$. This occurs because entanglement is not extracted in the limit of $\kappa \rightarrow 0$, where 
the two detectors follow comoving inertial worldlines and the distance between them is infinite \cite{KogaKM18}. 

Although we see that entanglement is extracted into the two detectors in this case, quantum teleportation is found to be impossible if these detectors are kept accelerated eternally, 
since classical communications between Alice and Bob, which is a necessary process in quantum teleportation, is forbidden in this case 
due to causality. However, this does not mean that the entanglement extracted this way is not usable in quantum teleportation. 
Indeed, if the detectors cease to be accelerated at a sufficiently late time and the causal contact between them is recovered, 
the standard quantum teleportation is possible using the entanglement extracted until that instant.

\subsubsection{Longitudinal shift} 

Next, we assume that Bob follows the worldline accelerated anti-parallelly to Alice, which is translated longitudinally, i.e., in the same 
spatial direction as the acceleration. For simplicity, we focus here on the case where the magnitudes of the accelerations of Alice and Bob 
are the same. We thus consider the worldline coordinates of Bob given as 
\begin{equation}
\bar{t}_B(\tau_B) = \frac{1}{\kappa} \sinh \left( \kappa \tau_B \right) , \quad 
\bar{x}_B(\tau_B) = - \frac{1}{\kappa} \cosh \left( \kappa \tau_B \right) + x_1 , \quad 
\bar{y}_B(\tau_B) = 0 , \quad \bar{z}_B(\tau_B) = 0 , 
\label{eqn:WorldLinesAntiAccelShift} 
\end{equation} 
and Alice's worldline coordinates are given by Eq. (\ref{eqn:WorldLineAlice}) with $\kappa_A = \kappa$. 
The constant $x_1$ is assumed to be non-zero, and constrained as 
\begin{equation}
\frac{2}{\kappa} > x_1 , 
\label{eqn:CondNonIntsctAntiAccelLg} 
\end{equation}
such that the two worldlines do not intersect. 
\begin{figure} 
\includegraphics[scale=0.5,keepaspectratio]{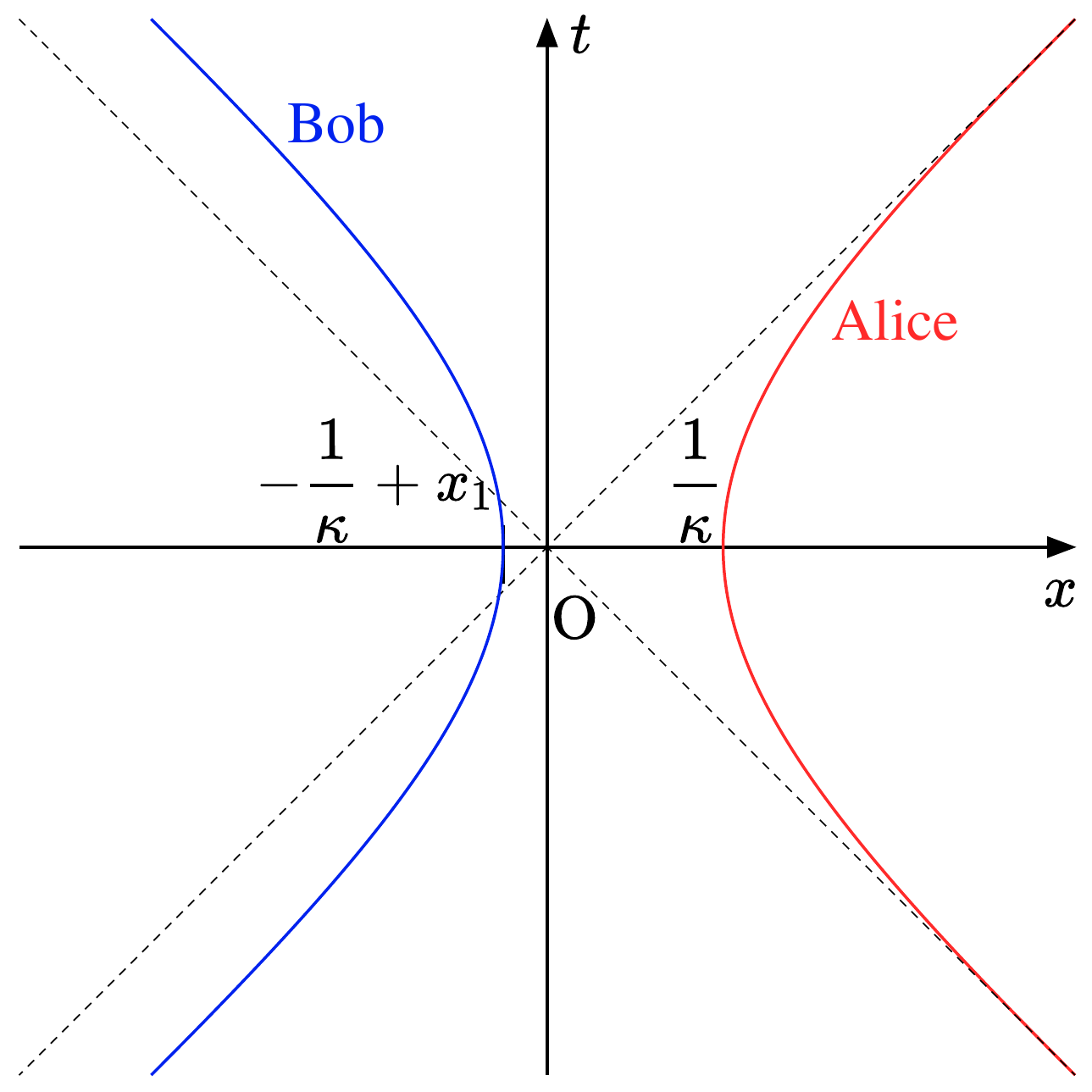} 
\caption{The worldlines of Alice and Bob accelerated anti-parallelly with a longitudinal shift to the acceleration.} 
\end{figure} 
The spacetime interval is described by Eq. (\ref{eqn:CausalityGen}) with 
\begin{equation} 
A(\tau_B) \equiv e^{\kappa \tau_B} - \kappa \, x_1  , \qquad 
B(\tau_B) \equiv \kappa \, x_1 \cosh \left( \kappa \tau_B \right) - \frac{\kappa^2 x_1^2}{2} - 1 , \qquad 
C(\tau_B) \equiv e^{- \kappa \tau_B} - \kappa \, x_1  , \qquad 
K \equiv \frac{- 1}{\kappa^2} , 
\label{eqn:ABCKAntiAccelLg} 
\end{equation} 
and then we obtain 
\begin{equation} 
D(\tau_B) 
= \frac{1}{2} \, \kappa \, x_1 \, e^{- \kappa \tau_B} 
\left( e^{\kappa \tau_B} - \zeta_+ \right) \left( e^{\kappa \tau_B} - \zeta_- \right) , 
\end{equation} 
where 
\begin{equation} 
\zeta_{\pm} 
= \frac{\kappa \, x_1}{2}  
\pm \sqrt{\frac{\kappa^2 x_1^2}{4} - 1}  . 
\end{equation} 

When $x_1$ is negative, we obtain 
\begin{align} & 
\tau_{A+n}(\tau_B) = \frac{1}{\kappa} \ln \left( e^{- \kappa \tau_B} - \kappa x_1 \right) 
+ \frac{1}{\kappa} ( 2 n + 1 ) \pi \ImUnit + \ImUnit \varepsilon , 
\quad 
\tau_{A-n}(\tau_B) = - \frac{1}{\kappa} \ln \left( e^{\kappa \tau_B} - \kappa \, x_1 \right) 
+ \frac{1}{\kappa} ( 2 n + 1) \pi \ImUnit - \ImUnit \varepsilon ,  
\end{align} 
and Eq. (\ref{eqn:CalIEFirstIntGen}) gives 
\begin{align} & 
\mathcal{I}_E 
= \frac{- \ImUnit}{4 \pi x_1} \frac{1}{\sinh \left( \pi \frac{\Delta E^{(A)}}{\kappa} \right)} 
\int_{- \infty}^{\infty} d \tau_B \, 
\frac{e^{\ImUnit \, \Delta E^{(B)} \tau_B} \, e^{\kappa \tau_B} }{
\left( e^{\kappa \tau_B} - \zeta_+ \right) \left( e^{\kappa \tau_B} - \zeta_- \right)} 
\left\{ e^{\ImUnit \frac{\Delta E^{(A)}}{\kappa} \ln \left( e^{- \kappa \tau_B} - \kappa  x_1 \right)} 
- e^{- \ImUnit \frac{\Delta E^{(A)}}{\kappa} \ln \left( e^{\kappa \tau_B} - \kappa x_1 \right)} \right\} . 
\end{align} 
The triangle inequality and the integral inequality give 
\begin{equation} 
\left| \mathcal{I}_E \right|  \leq 
\frac{1}{2 \pi \kappa \left| x_1 \right|} \frac{1}{\sinh \left( \pi \frac{\Delta E^{(A)}}{\kappa} \right)}  \, P , 
\end{equation} 
where 
\begin{equation} 
P 
= \left\{ \begin{array}{lll} 
\dfrac{1}{\sqrt{1 - \dfrac{\kappa^2 x_1^2}{4}}} \, \arctan \dfrac{2 \sqrt{1 - \dfrac{\kappa^2 x_1^2}{4}}}{\kappa \left| x_1 \right|} 
& \quad \mathrm{for} \quad & 1 > \dfrac{\kappa^2 x_1^2}{4}  \\ 
\dfrac{1}{\sqrt{\dfrac{\kappa^2 x_1^2}{4} - 1}} \, \ln \left( \dfrac{\kappa \left| x_1 \right|}{2} + \sqrt{\dfrac{\kappa^2 x_1^2}{4} 
- 1} \right) 
& \quad \mathrm{for} \quad & 1 < \dfrac{\kappa^2 x_1^2}{4} \\ 
1 & \quad \mathrm{for} \quad & 1 = \dfrac{\kappa^2 x_1^2}{4} 
\end{array} \right. . 
\label{eqn:PDefAntiAccelLg} 
\end{equation} 
In the case of positive $x_1$, where $x_1$ is restricted as $0 < x_1 < \dfrac{2}{\kappa}$ from Eq. (\ref{eqn:CondNonIntsctAntiAccelLg}), 
we have 
\begin{align} & 
\tau_{A+n}(\tau_B) = \left\{ \begin{array}{lll} 
\dfrac{1}{\kappa} \ln \left( e^{- \kappa \tau_B} - \kappa \, x_1 \right) 
+ \dfrac{1}{\kappa} ( 2 n + 1 ) \pi \ImUnit - \ImUnit \varepsilon 
& \; & \tau_B < - \dfrac{1}{\kappa} \ln \kappa \, x_1 \\ \\ 
\dfrac{1}{\kappa} \ln \left( \kappa \, x_1  - e^{- \kappa \tau_B} \right) 
+ \dfrac{1}{\kappa} 2 n \pi \ImUnit - \ImUnit \varepsilon 
& \; & - \dfrac{1}{\kappa} \ln \kappa \, x_1 < \tau_B 
\end{array} \right. 
\notag  \\ & 
\tau_{A-n}(\tau_B) = \left\{ \begin{array}{lll} 
- \dfrac{1}{\kappa} \ln \left( \kappa x_1 - e^{\kappa \tau_B} \right) 
+ \dfrac{1}{\kappa} 2 n \pi \ImUnit + \ImUnit \varepsilon 
& \; & \tau_B < \dfrac{1}{\kappa} \ln \kappa \, x_1 \\ \\ 
- \dfrac{1}{\kappa} \ln \left( e^{\kappa \tau_B} - \kappa \, x_1 \right) 
+ \dfrac{1}{\kappa} ( 2 n + 1 ) \pi \ImUnit + \ImUnit \varepsilon 
& \; & \dfrac{1}{\kappa} \ln \kappa \, x_1 < \tau_B 
\end{array} \right. , 
\end{align} 
and we obtain from Eq. (\ref{eqn:CalIEFirstIntGen}) 
\begin{align} & 
\mathcal{I}_E 
= \frac{- \ImUnit}{2 \pi x_1} \:  \frac{1}{1 - e^{- 2 \pi \frac{\Delta E^{(A)}}{\kappa}}} 
\Bigg[ 
e^{- \pi \frac{\Delta E^{(A)}}{\kappa}} \, \int_{- \infty}^{- \frac{1}{\kappa} \ln \kappa x_1} d \tau_B \, 
\frac{e^{\ImUnit \, \Delta E^{(B)} \tau_B} \, e^{\kappa \tau_B} }{
\left( e^{\kappa \tau_B} - \zeta_+ \right) \left( e^{\kappa \tau_B} - \zeta_- \right)} 
e^{\ImUnit \frac{\Delta E^{(A)}}{\kappa} \ln \left( e^{- \kappa \tau_B} - \kappa x_1  \right)} 
\notag \\ & 
+ e^{- 2 \pi \frac{\Delta E^{(A)}}{\kappa}} \, \int_{- \frac{1}{\kappa} \ln \kappa x_1}^{\infty} d \tau_B \, 
\frac{e^{\ImUnit \, \Delta E^{(B)} \tau_B} \, e^{\kappa \tau_B} }{
\left( e^{\kappa \tau_B} - \zeta_+ \right) \left( e^{\kappa \tau_B} - \zeta_- \right)} 
e^{\ImUnit \frac{\Delta E^{(A)}}{\kappa} \ln \left( \kappa x_1 - e^{- \kappa \tau_B} \right)} 
\notag \\ & 
- \int_{- \infty}^{\frac{1}{\kappa} \ln \kappa x_1} d \tau_B \, 
\frac{e^{\ImUnit \, \Delta E^{(B)} \tau_B} \, e^{\kappa \tau_B} }{
\left( e^{\kappa \tau_B} - \zeta_+ \right) \left( e^{\kappa \tau_B} - \zeta_- \right)} 
e^{- \ImUnit \frac{\Delta E^{(A)}}{\kappa} \ln \left( \kappa x_1 - e^{\kappa \tau_B} \right)}  
\notag \\ & 
- e^{- \pi \frac{\Delta E^{(A)}}{\kappa}} \, \int_{\frac{1}{\kappa} \ln \kappa x_1}^{\infty} d \tau_B \, 
\frac{e^{\ImUnit \, \Delta E^{(B)} \tau_B} \, e^{\kappa \tau_B} }{
\left( e^{\kappa \tau_B} - \zeta_+ \right) \left( e^{\kappa \tau_B} - \zeta_- \right)} 
e^{- \ImUnit \frac{\Delta E^{(A)}}{\kappa} \ln \left( e^{\kappa \tau_B} - \kappa x_1 \right)}  
\Bigg] . 
\end{align} 
A similar computation as in the case of $x_1 < 0$ leads to 
\begin{equation} 
\left| \mathcal{I}_E \right| \leq 
\frac{1}{\pi \kappa x_1} \:  \frac{1}{1 - e^{- 2 \pi \frac{\Delta E^{(A)}}{\kappa}}} \, P  , 
\end{equation} 
where $P$ is defined by Eq. (\ref{eqn:PDefAntiAccelLg}), but only the case of $1 > \kappa^2 x_1^2 / 4$ is possible here. 
Therefore, in any case, $\left| \mathcal{I}_E \right|$ is bounded. 
As in the parallel acceleration, longitudinally shifted worldlines are associated with different timelike boost Killing vector fields, and thus  
$\mathcal{I}_E$ may not vanish. However, it is dominated by 
the excitation probability $\mathcal{I}_I$ at a sufficiently late time, and thus it is impossible to extract entanglement from the vacuum into the two detectors, again. 

%
\subsection{Acceleration in oriented directions} 

We consider finally the case where Alice and Bob are accelerated with the same magnitude $\kappa$ of the acceleration, 
but in the directions orientated differently with respect to each other. The worldline of Alice is thus given by Eq. (\ref{eqn:WorldLineAlice}) with 
$\kappa_A = \kappa$, and Bob's worldline coordinates are given by 
\begin{equation}
\bar{t}_B(\tau_B) = \frac{1}{\kappa} \, \sinh \left( \kappa \tau_B \right) , \quad 
\bar{x}_B(\tau_B) = \frac{1}{\kappa} \, \cosh \left( \kappa \tau_B \right) \cos \phi , \quad
\bar{y}_B(\tau_B) = \frac{1}{\kappa} \, \cosh \left( \kappa \tau_B \right) \sin \phi , \quad
\bar{z}_B(\tau_B) = 0 , 
\label{eqn:WorldlinesAccelRot} 
\end{equation} 
where we restrict the range of the angle $\phi$ between the two worldlines, as 
\begin{equation} 
0 < \phi < \pi . 
\label{eqn:PhiRangeRotAccel} 
\end{equation} 
The case of $\phi = 0$ implies that Bob adheres to Alice all through the time, and hence the two detectors will be 
naturally entangled in this case. We are thus not concerned with the case of $\phi = 0$. On the other hand, the case of 
$\phi = \pi$ corresponds to the particular case of $\sigma = 0$ in Sec \ref{sec:Entangled}, which we do not repeat here. 
\begin{figure} 
\includegraphics[scale=0.5,keepaspectratio]{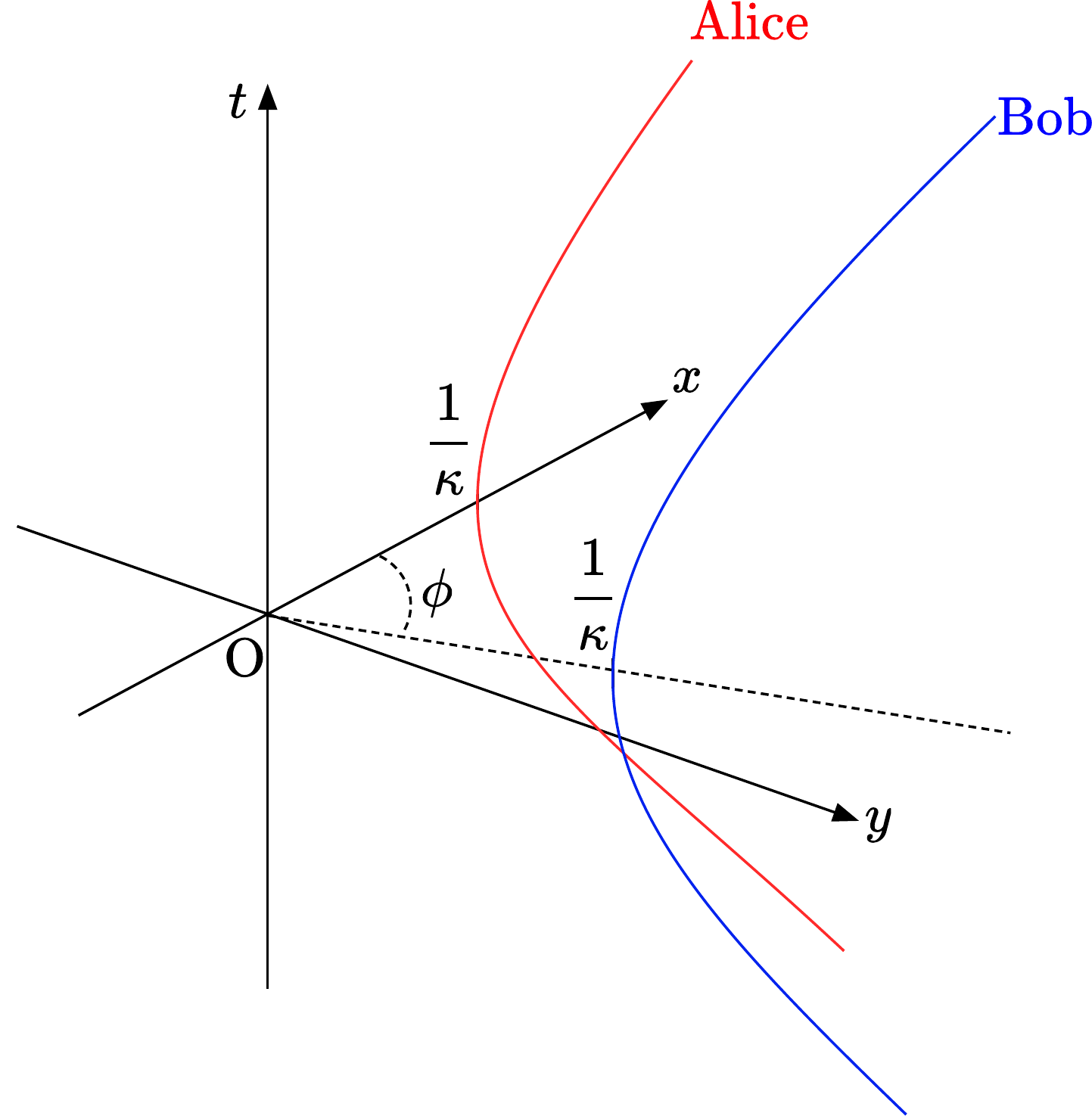} 
\caption{The worldlines of Alice and Bob accelerated in orientated directions with respect to each other.} 
\end{figure} 
The spacetime interval in the present case is described by Eq. (\ref{eqn:CausalityGen}) with 
\begin{align} & 
A(\tau_B) \equiv - \sin^2 \frac{\phi}{2} e^{- \kappa \tau_B} 
\left( e^{\kappa \tau_B} - \cot \frac{\phi}{2} \right) \left( e^{\kappa \tau_B} + \cot \frac{\phi}{2} \right)  , \qquad 
B(\tau_B) \equiv 1 , 
\label{eqn:ABDefRotAccel} \\ & 
C(\tau_B) \equiv \cos^2 \frac{\phi}{2} e^{- \kappa \tau_B} 
\left( e^{\kappa \tau_B} - \tan \frac{\phi}{2} \right) \left( e^{\kappa \tau_B} + \tan \frac{\phi}{2} \right)  , \qquad 
K \equiv \frac{1}{\kappa^2} , 
\label{eqn:KDefRotAccel} 
\end{align} 
and we derive as 
\begin{equation} 
D(\tau_B) = \sin \phi \, \cosh \left( \kappa \tau_B \right) , 
\label{eqn:DDefRotAccel} 
\end{equation}  
along with 
\begin{align} & 
\tau_{A+n}(\tau_B) = \left\{ \begin{array}{lll} 
T_+(\tau_B) + \dfrac{1}{\kappa} 2 n \pi \ImUnit + \ImUnit \varepsilon 
& \quad & \tau_B < \dfrac{1}{\kappa} \ln \cot \dfrac{\phi}{2} \\ \\ 
T_+(\tau_B) + \dfrac{1}{\kappa} ( 2 n + 1 ) \pi \ImUnit + \ImUnit \varepsilon 
& \quad & \tau_B > \dfrac{1}{\kappa} \ln \cot \dfrac{\phi}{2}
\end{array} \right. , 
\label{eqn:TauAPNRotAccel} \\ & 
\tau_{A-n}(\tau_B) = \left\{ \begin{array}{lll} 
T_-(\tau_B) + \dfrac{1}{\kappa} ( 2 n + 1 ) \pi \ImUnit - \ImUnit \varepsilon  
& \quad & \tau_B < \dfrac{1}{\kappa} \ln \tan \dfrac{\phi}{2} \\ \\ 
T_-(\tau_B) + \dfrac{1}{\kappa} 2 n \pi \ImUnit - \ImUnit \varepsilon  
& \quad &  \tau_B > \dfrac{1}{\kappa} \ln \tan \dfrac{\phi}{2}
\end{array} \right. , 
\label{eqn:TauAMNRotAccel}
\end{align} 
where the real functions $T_{\pm}(\tau_B)$ are defined as 
\begin{equation}
T_+(\tau_B) \equiv \dfrac{1}{\kappa} \ln \left| \cot \dfrac{\phi}{2} \, \dfrac{e^{\kappa \tau_B} + \tan \dfrac{\phi}{2}}{e^{\kappa \tau_B} - \cot \dfrac{\phi}{2}} \right| , \quad 
T_-(\tau_B) \equiv  \dfrac{1}{\kappa} \ln \left| \cot \dfrac{\phi}{2} \, \dfrac{e^{\kappa \tau_B} - \tan \dfrac{\phi}{2}}{e^{\kappa \tau_B} + \cot \dfrac{\phi}{2}} \right| .  
\label{eqn:TPMDefRotAccel} 
\end{equation} 
By substituting Eqs. (\ref{eqn:ABDefRotAccel})--(\ref{eqn:TPMDefRotAccel}) into Eq. (\ref{eqn:CalIEFirstIntGen}), we compute as 
\begin{align} & 
\mathcal{I}_E 
= \frac{\ImUnit \kappa}{4 \pi \sin \phi} \frac{1}{1 - e^{- 2 \pi \frac{\Delta E^{(A)}}{\kappa}}} \: 
\notag \\ & \qquad \times 
\left[  \int_{- \infty}^{\frac{1}{\kappa} \ln \cot \frac{\phi}{2}} d \tau_B \, 
\frac{e^{\ImUnit \, \Delta E^{(B)} \tau_B}}{\cosh \left( \kappa \tau_B \right)} 
e^{\ImUnit \Delta E^{(A)} T_+(\tau_B)} 
+ e^{- \pi \frac{\Delta E^{(A)}}{\kappa}} \, \int_{\frac{1}{\kappa} \ln \cot \frac{\phi}{2}}^{\infty} d \tau_B \, 
\frac{e^{\ImUnit \, \Delta E^{(B)} \tau_B}}{\cosh \left( \kappa \tau_B \right)} 
e^{\ImUnit \Delta E^{(A)} T_+(\tau_B)} 
\right. \notag \\ & \left. 
-  e^{- \pi \frac{\Delta E^{(A)}}{\kappa}} \, \int_{- \infty}^{\frac{1}{\kappa} \ln \tan \frac{\phi}{2}} d \tau_B \, 
\frac{e^{\ImUnit \, \Delta E^{(B)} \tau_B}}{\cosh \left( \kappa \tau_B \right)} 
e^{\ImUnit \Delta E^{(A)} T_-(\tau_B)} 
-  e^{- 2 \pi \frac{\Delta E^{(A)}}{\kappa}} \, \int_{\frac{1}{\kappa} \ln \tan \frac{\phi}{2}}^{\infty} d \tau_B \, 
\frac{e^{\ImUnit \, \Delta E^{(B)} \tau_B}}{\cosh \left( \kappa \tau_B \right)} 
e^{\ImUnit \Delta E^{(A)} T_-(\tau_B)} 
\right]  . 
\label{eqn:CalIERotAccel} 
\end{align} 
We then resort to the triangle inequality and the integral inequality and obtain 
\begin{equation}
\left| \mathcal{I}_E \right| 
\leq \frac{1}{4 \sin \phi} \coth \left( \pi \frac{\Delta E^{(A)}}{\kappa} \right) \left[ 
1- \frac{\phi}{\pi} \tanh \left( \pi \frac{\Delta E^{(A)}}{\kappa} \right)\tanh \left( \frac{\pi}{2} \frac{\Delta E^{(A)}}{\kappa} \right) \right] . 
\label{eqn:CalIEBndRotAccel} 
\end{equation}
Thus, we see that $\left| \mathcal{I}_E \right|$ is bounded within the range (\ref{eqn:PhiRangeRotAccel}) of $\phi$, 
and therefore the two detectors are not entangled.  
Clearly, the boost Killing vector fields of Alice and Bob are different from each other in this case. 
This will be the reason why the delta function do not appear and thus $\mathcal{I}_E$ may not vanish, as above.

\section{Summary and discussion} 
\label{sec:summary} 

We considered a pair of two-level Unruh-DeWitt detectors, both of which are uniformly accelerated, in the Minkowski 
vacuum of a massless neutral scalar field. The worldlines of parallelly accelerated, anti-parallelly accelerated, and 
accelerated in differently orientated directions are considered with a translational shift and/or different magnitudes of the acceleration. 
The initial state of the whole system at the asymptotic past is assumed to be the ground state, 
and we computed within the standard perturbation theory the entanglement between the two 
detectors at the asymptotic future. 
We presented the single framework where these cases are analyzed in a unified manner. 
Although this framework is applicable to numerical computations so long as one of the detectors follows a uniformly accelerated 
worldline,  we focus on the analytically tractable cases so that we can manipulate the delta function associated with 
the energy conservation. 

In the case of parallel acceleration, we considered the cases with a transverse shift, different magnitudes of acceleration, and a longitudinal shift. 
In any of these cases, we showed that the two detectors are not entangled at the asymptotic future. 
In the case of anti-parallel acceleration, we considered the case with a transverse shift along with different magnitudes of the acceleration, 
and found that entanglement can be extracted into the detectors only when the ratios $\Delta E^{(I)} / \kappa_I$ coincide between the two detectors. 
We found that although the parameter space for entanglement extraction increases as the ratio $\Delta E^{(I)} / \kappa_I$ gets larger, 
the amount of the entanglement actually decreases, which is in accordance with the result in the case of inertial motions \cite{KogaKM18}. 
We also considered the case of anti-parallel acceleration with a longitudinal shift, but we saw that the detectors are not entangled in this case. 
When the detectors are accelerated in differently orientated directions with each other, we showed that entanglement is not extracted from the vacuum, 
in contrast to the recent paper \cite{GrochowskiLD19-}, where the entanglement of two wave packets at some instant is considered. 
Thus, as long as a pair of uniformly accelerated Unruh-DeWitt detectors are concerned, the standard quantum teleportation is not possible. 
In the case of causally connected detectors, entanglement is not extracted, while classical communication, which is necessary 
in quantum teleportation, is forbidden when entanglement is extracted. 

We discussed these results from the viewpoint of the timelike boost Killing vector fields tangent to the worldlines of the detectors. 
We saw that if the timelike boost Killing vector fields 
coincide between Alice and Bob, the energy conservation leads to the expression of $\mathcal{I}_E$ involving the delta function. 
In the case of parallel acceleration, this forces $\mathcal{I}_E$ in Eq. (\ref{eqn:CalIEParaAccel}) to vanish, because we assumed the excitation energy $\Delta E^{(I)}$ is larger than zero. 
On the other hand, in the case of anti-parallel acceleration, the very same fact makes the detectors entangled, as shown 
in Eq. (\ref{eqn:CalIEResAntiAccelTr}), and we pointed out that this results from the behavior of the Rindler modes. 
In the case when the timelike boost Killing vector fields do not coincide between the detectors, we found that the delta function does not appear in 
the expression of $\mathcal{I}_E$ and it may not vanish. However, even in the latter case, the excitation probability $\mathcal{P}_I$ 
due to thermal fluctuations dominates the correlation described by $\mathcal{I}_E$, and then entanglement is not extracted. 

This argument based on the timelike Killing vector field applies to the case of inertial motions \cite{KogaKM18}, where 
the detectors are not entangled if they are comoving, while they are entangled if they are in a relative motion. In the latter case, 
the timelike Killing vector field along the worldline of one of the detectors is a linear combination of the timelike Killing vector of the other detector and the spatial translational Killing vector. Therefore, the energy of a virtual quantum emitted from one detector is not necessarily the same as 
the one absorbed by the other detector, which results in non-vanishing $\mathcal{I}_E$ and hence gives rise to entanglement in the case of 
relative inertial motions. 
However, in the case of uniformly accelerated motions, thermal noise overcomes this effect. 
Indeed, based on the non-perturbative dynamical analysis \cite{LinHu10}, we expect that the entanglement will be degraded 
due to the thermal effect, 
if the coupling between the detectors and the scalar field lasts long enough that the perturbation theory breaks down and 
the thermal equilibrium is achieved between the detectors and the scalar field. 

However, it does not spoil the perturbative investigation in this paper. 
Actually, by switching on and off the detectors at the asymptotic past and future but within the regime where the perturbation theory 
is valid, one can extract entanglement from the Minkowski vacuum into Unruh-DeWitt detectors uniformly accelerated anti-parallelly. 
Rather, the perturbative analysis may be helpful enough in order to {\it probe} the features of entanglement contained in the vacuum, 
without drastically changing the quantum state to be probed. 

As in the case of inertial motions \cite{KogaKM18}, the relevance of the energy conservation in entanglement extraction 
will be understood as resulting from the suppression of the uncertainty between time and energy, due to a infinitely long interaction time. 
Although energy is imparted to the system under consideration in the case of uniformly accelerated detectors, 
we found that the energy conservation still plays one of decisive roles in entanglement extraction from the vacuum. 
We thus expect that the result in this paper has uncovered a fundamental aspect of entanglement extraction from the vacuum, 
which in turn will be closely related with the nature of the entanglement contained in the vacuum. 
In addition, based on our result, one can discuss other physical effects in entanglement extraction, 
such as the spatial extension and the structure of detectors, the spatial profile and the dynamics of wave packets, and quantum fluctuations due to a restricted period of interaction. 
We expect also that the role of the energy conservation in entanglement extraction might provide a new insight into the relation between 
entanglement and energy, such as those in Refs.  \cite{BhattacharyaNTU13,BenyCFO18}.

\acknowledgments 
This work was supported in part by JSPS KAKENHI Grant Number 17K18107 and 17K05451. 

\appendix 

\section{Uniformly accelerated detectors with relative velocity and inertial limit} 

In this appendix, we consider the case where Alice and Bob follow worldlines with uniform acceleration, which reduce to those of 
an inertial relative motion in the vanishing acceleration limit, and we will see that it reproduces the result presented in Ref. \cite{KogaKM18}. 

For this purpose, we assume that the worldline coordinates of Alice are given as 
\begin{equation}
\bar{t}_A(\tau_A) = \frac{1}{\kappa} \sinh \left( \kappa \, \tau_A \right) , \quad 
\bar{x}_A(\tau_A) = \frac{1}{\kappa} \left[ \cosh \left( \kappa \, \tau_A \right) - 1 \right] , \quad 
\bar{y}_A(\tau_A) = 0 , \quad \bar{z}_A(\tau_A) = 0 , 
\label{eqn:AliceWLAccelBoost} 
\end{equation}  
and those of Bob as 
\begin{align} & 
\bar{t}_B(\tau_B) 
= \frac{1}{\kappa} \left[ \sinh \left( \kappa \, \tau_B + \alpha \right) - \sinh \alpha  \right]  , \quad 
\bar{x}_B(\tau_B) 
=  \frac{1}{\kappa} \left[ \cosh \left( \kappa \, \tau_B + \alpha \right) - \cosh \alpha \right] , \quad  
\notag \\ & 
\bar{y}_B(\tau_B) = y_0 , \quad \bar{z}_A(\tau_A) = z_0 , 
\label{eqn:BobWLAccelBoost} 
\end{align} 
where $\alpha$, $y_0$, and $z_0$ are arbitrary non-vanishing constants, and $\kappa$ is the magnitude of the acceleration. 
In the limit of $\kappa \rightarrow 0$, Eqs. (\ref{eqn:AliceWLAccelBoost}) and (\ref{eqn:BobWLAccelBoost}) reduce to 
the worldline coordinates of inertial relative motions with the relative velocity $v = \tanh \alpha$ as 
\begin{align} & 
\bar{t}_A(\tau_A) = \tau_A  , \quad  \bar{x}_A(\tau_A) = 0 , \quad \bar{y}_A(\tau_A) = 0 , \quad \bar{z}_A(\tau_A) = 0 , 
\notag \\ & 
\bar{t}_B(\tau_B) = \frac{1}{\sqrt{1 - v^2}} \tau_B  , \quad 
\bar{x}_B(\tau_B) =   \frac{v}{\sqrt{1 - v^2}} \tau_B , \quad  
\bar{y}_B(\tau_B) = y_0 , \quad \bar{z}_A(\tau_A) = z_0 . 
\end{align} 

In this case, the constant $K$ and  the functions $A(\tau_B)$, $B(\tau_B)$, and $C(\tau_B)$ in the spacetime interval Eq. (\ref{eqn:CausalityGen}) are found to be given as 
\begin{align} & 
K \equiv \frac{1}{\kappa^2} , 
\qquad 
A(\tau_B) \equiv \left( e^{-  \kappa \, \tau_B } - 1 \right)  e^{- \alpha} + 1 , 
\label{eqn:ADefBoostAccel} \\ & 
B(\tau_B) \equiv 1 + \frac{\kappa^2 \varrho_0^2}{2} 
+ 4  \cosh \frac{\kappa \, \tau_B + \alpha}{2} \sinh \frac{\kappa \tau_B}{2} \sinh \frac{\alpha}{2} , 
\label{eqn:BDefBoostAccel} \\ & 
C(\tau_B) \equiv \left( e^{\kappa \, \tau_B} - 1 \right) e^{\alpha} + 1 , 
\label{eqn:CDefBoostAccel} 
\end{align} 
where $\varrho_0$ is defined by Eq. (\ref{eqn:Rho0Def}). 
The functions $D(\tau_B)$ defined in Eq. (\ref{eqn:DDefGen}) and $\tau_{A\pm n}(\tau_B)$ defined in Eq. (\ref{eqn:TauAPMNDefGen}) are 
derived as 
\begin{align} & 
D(\tau_B) 
= \sqrt{\kappa^2 \varrho_0^2 
+ \left( \frac{\kappa^2 \varrho_0^2}{2} + 4  \cosh \frac{\kappa \, \tau_B + \alpha}{2} \sinh \frac{\kappa \tau_B}{2} \sinh \frac{\alpha}{2} \right)^2} , 
\label{eqn:DDefAccelBoost} 
\end{align} 
and 
\begin{align} & 
\tau_{A+n}(\tau_B) = \left\{ \begin{array}{lll} 
\dfrac{1}{\kappa} \ln \Upsilon_+(\tau_B) + \dfrac{1}{\kappa} 2 n \pi \ImUnit +  \ImUnit \varepsilon 
& \quad & \Upsilon_+(\tau_B) > 0 \\ \\
\dfrac{1}{\kappa} \ln \left| \Upsilon_+(\tau_B) \right| + \dfrac{1}{\kappa} ( 2 n + 1 ) \pi \ImUnit +  \ImUnit \varepsilon 
& \quad & \Upsilon_+(\tau_B) < 0 
\end{array} \right. , 
\notag \\ & 
\tau_{A-n}(\tau_B) = \left\{ \begin{array}{lll} 
\dfrac{1}{\kappa} \ln \Upsilon_-(\tau_B) + \dfrac{1}{\kappa} 2 n \pi \ImUnit -  \ImUnit \varepsilon 
& \quad & \Upsilon_-(\tau_B) > 0 \\ \\ 
\dfrac{1}{\kappa} \ln \left| \Upsilon_-(\tau_B) \right| + \dfrac{1}{\kappa} ( 2 n + 1 ) \pi \ImUnit -  \ImUnit \varepsilon 
& \quad & \Upsilon_-(\tau_B) < 0 
\end{array} \right. ,  
\label{eqn:TauPMnDefAccelBoost} 
\end{align} 
where $\Upsilon_{\pm}(\tau_B) \equiv \left[ B(\tau_B) \pm D(\tau_B) \right] / A(\tau_B)$. 

By substituting Eqs. (\ref{eqn:DDefAccelBoost}) and (\ref{eqn:TauPMnDefAccelBoost}) into Eq. (\ref{eqn:CalIEFirstIntGen}) with 
$K = 1 / \kappa^2$ and $\kappa_A = \kappa$, we obtain 
\begin{align} & 
\mathcal{I}_E =  \frac{\ImUnit \kappa}{4 \pi} \: \frac{1}{1 - e^{- 2 \pi  \frac{\Delta E^{(A)}}{\kappa}}} \:  \int_{- \infty}^{\infty} d \tau_B \, 
\frac{e^{\ImUnit \, \Delta E^{(B)} \tau_B}}{D(\tau_B)}  
\notag \\ & \times 
\left( e^{\ImUnit \frac{\Delta E^{(A)}}{\kappa} \ln \left| \Upsilon_+(\tau_B) \right|} 
\left[ \frac{1 + \mathrm{sgn} \left( \Upsilon_+(\tau_B) \right)}{2} + \frac{1 - \mathrm{sgn} \left( \Upsilon_+(\tau_B) \right)}{2} e^{- \pi  \frac{\Delta E^{(A)}}{\kappa}} \right] 
\right. \notag \\ & \left. 
- e^{- \pi  \frac{\Delta E^{(A)}}{\kappa}}  \: 
e^{\ImUnit \frac{\Delta E^{(A)}}{\kappa} \ln \left| \Upsilon_-(\tau_B) \right|} 
\left[ \frac{1 + \mathrm{sgn} \left( \Upsilon_-(\tau_B) \right)}{2} e^{- \pi  \frac{\Delta E^{(A)}}{\kappa}} + \frac{1 - \mathrm{sgn} \left( \Upsilon_-(\tau_B) \right)}{2} \right] \right) . 
\label{eqn:CalEGFAlAcBobAcBoostFin}
\end{align} 
Although numerical work is necessary to compute Eq. (\ref{eqn:CalEGFAlAcBobAcBoostFin}) in general, our interest here is 
the limit $\kappa \rightarrow 0$. To leading order in $\kappa$, Eq. (\ref{eqn:CalEGFAlAcBobAcBoostFin}) is approximated as 
\begin{align} & 
\mathcal{I}_E =  \frac{\ImUnit}{4 \pi} \: \frac{\sqrt{1 - v^2}}{v}  \int_{- \infty}^{\infty} d \tau_B \, 
\frac{1}{\sqrt{\tau_B^2 + \ell^2}}  \: 
e^{\ImUnit \, \epsilon \, \tau_B} \, 
e^{\ImUnit \, p \,  \sqrt{\tau_B^2 + \ell^2}} 
=  \frac{\ImUnit}{2 \pi} \: \frac{\sqrt{1 - v^2}}{v} \, K_0(\ell \sqrt{\epsilon^2 - p^2}) .
\end{align} 
where $K_0$ is the zeroth modified Bessel function of the second kind and 
\begin{equation}
\ell \equiv \frac{\sqrt{1 - v^2}}{v} \varrho_0 
, \quad 
\epsilon \equiv \Delta E^{(B)} + \frac{\Delta E^{(A)}}{\sqrt{1 - v^2}} 
, \quad 
p \equiv \frac{v}{\sqrt{1 - v^2}} \Delta E^{(A)} . 
\end{equation} 
On the other hand, $\mathcal{I}_I$ is exponentially suppressed as $1/\kappa$ grows, and hence its contribution is negligible in the limit of $\kappa \rightarrow 0$.  
Therefore, we see that we have reproduced the result in the case of an inertial relative motion presented in Ref. \cite{KogaKM18},  
based on the framework in this paper. 

%
\vskip 2cm
\baselineskip .2in
 
\end{document}